\documentclass[numbers=endperiod]{scrartcl}
\usepackage{ebgaramond}

\usepackage{graphicx}

\usepackage{float}
\usepackage{amsmath}
\usepackage{multirow}
\usepackage{breakcites}
\usepackage{enumitem}
\usepackage{hyperref}
\usepackage{authblk}
\usepackage[symbol]{footmisc}
\usepackage{wrapfig}
\usepackage{caption}

\DeclareCaptionFormat{myformat}{\makebox[0.5in][l]{#1}\parbox[t]{\dimexpr \textwidth - 0.5in}{#3}}

\usepackage{subcaption}
\usepackage{cleveref}
\usepackage{titlesec}

\titleformat{\section}
  {\normalfont\scshape\Large\bfseries}
  {\thesection}{1em}{}
\titleformat{\subsection}
  {\normalfont\scshape\large\bfseries}
  {\thesubsection}{1em}{}
 \titleformat{\subsubsection}
  {\normalfont\scshape\bfseries}
  {\thesubsubsection}{1em}{}

\setkomafont{title}{\normalfont\huge\scshape}
\setkomafont{subtitle}{\normalfont\LARGE\scshape}

\usepackage[english]{babel}
\addto\captionsenglish{}

\usepackage{lineno}
\usepackage{setspace}
\title{\textsc{Polarize, Catalyze, Stabilize}}
\subtitle{How a minority of norm internalizers amplifies group selection and punishment}

\begin{document}
\title{\textsc{Polarize, Catalyze, Stabilize}}
\subtitle{How a minority of norm internalizers amplify group selection and punishment}
\author[1*]{Victor Vikram Odouard}
\author[2]{Diana Smirnova}
\author[3]{Shimon Edelman}
\affil[1]{\begin{normalsize}Santa Fe Institute\end{normalsize}}
\affil[2]{\begin{normalsize}Cornell University, Department of Computer Science\end{normalsize}}
\affil[3]{\begin{normalsize}Cornell University, Department of Psychology\end{normalsize}}

\begin{center}
    \makeatletter
    \begin{huge}
        \@title \\[0.2em]
    \end{huge}
    \begin{LARGE}
        \@subtitle \\[0.8em]
    \end{LARGE}
    \begin{Large}
        \emph{by}\\
      \@author \\[2em]
    \end{Large}
\end{center}

\begin{footnotesize} \noindent * Author to whom correspondence should be addressed: Victor Vikram Odouard, Santa Fe Institute (Cowan Campus) 1399 Hyde Park Road, Santa Fe, New Mexico 87501, U.S.A. E-mail: vo47@cornell.edu\end{footnotesize}

\begin{center}
    \textsc{Abstract}
\end{center}
Many mechanisms behind the evolution of cooperation, such as reciprocity, indirect reciprocity, and altruistic punishment, require group knowledge of individual actions. But what keeps people cooperating when no one is looking? Conformist norm internalization, the tendency to abide by the behavior of the majority of the group, even when it is individuallly harmful, could be the answer. In this paper, we analyze a world where (1) there is group selection and punishment by indirect reciprocity but (2) many actions (half) go unobserved, and therefore unpunished. Can norm internalization fill this `observation gap' and lead to high levels of cooperation, even when agents may in principle cooperate only when likely to be caught and punished? Specifically, we seek to understand whether adding norm internalization to the strategy space in a public goods game can lead to higher levels of cooperation when both norm internalization and cooperation start out rare. We found the answer to be positive, but, interestingly, not because norm internalizers end up making up a substantial fraction of the population, nor because they cooperate much more than other agent types. Instead, norm internalizers, by polarizing, catalyzing, and stabilizing cooperation, can increase levels of cooperation of other agent types, while only making up a minority of the population themselves.

\begin{center}
    \textsc{Significance Statement}
\end{center}
What keeps people cooperating when no one is looking? Group selection favoring cooperative groups does not require observers, but often works best when there is some other norm enforcement mechanism to supplement it. But most norm enforcement mechanisms require observers in order to function. Here we examine whether norm internalization could fill this enforcement gap, by acting as an “internal enforcer” of norms. Perhaps surprisingly, we discovered in our simulations that the population of norm internalizers always remained quite small, and norm internalizers didn’t necessarily cooperate more than other agent types. Nevertheless, under certain conditions, they were able to spark significantly higher mean levels of cooperation, by causing other agents to cooperate more – partly by sparking bouts of high cooperation after periods of very low cooperation, and partly by helping maintain those bouts of high cooperation for longer. 

\begin{center}
    \textsc{Keywords}

\emph{norm internalization $\cdot$ evolution $\cdot$ group selection $\cdot$ cooperation $\cdot$ public goods game $\cdot$ conformist transmission}
\end{center}

\begin{center}
    \textsc{Highlights}
\begin{itemize}
    \item in conditions of group selection and punishment, with both cooperation and norm internalization starting rare, norm internalization only spreads to a minority of the population
    \item norm internalization is nonetheless able to increase mean levels of cooperation, mostly by increasing the cooperation of other agent types.
    \item norm internalization plays three roles: it polarizes groups to extremes, it catalyzes cooperation spikes when cooperation starts especially low, and it stabilizes high levels of cooperation
\end{itemize}
\end{center}

\section{Introduction}
Because altruism, as expressed in unconditional cooperation or in self-sacrifice, is by itself not an evolutionarily stable trait, its persistence requires a special explanation \cite{darwinDescentManVolume2011}. Almost all such explanations share the characteristic that they \emph{make altruists more likely to accrue the benefits of altruism from others}. 

In kin altruism the agent benefits from the altruism of its relatives, who are disproportionately likely to share its altruistic gene \cite{hamiltonGeneticalEvolutionSocial1964, quellerGeneralModelKin1992}. Reciprocity ensures that the agent benefits from the help of those it helped in the past, either directly \cite{triversEvolutionReciprocalAltruism1971, axelrodEmergenceCooperationEgoists1981, axelrodEvolutionCooperation1981, brownEvolutionSocialBehavior1982}, or indirectly, when it benefits from those who act on its reputation as a cooperator \cite{panchanathanTaleTwoDefectors2003, nowakEvolutionIndirectReciprocity2005}. Because punishers coerce those around them into cooperating, they, too, are more likely to accrue benefits of altruism \cite{boydPunishmentAllowsEvolution1992, boydEvolutionAltruisticPunishment2003}. Finally, under group selection altruistic groups are more likely to proliferate, causing the majority of altruists to reside in highly altruistic groups and therefore benefitting collectively from altruism \cite{bowlesCoevolutionIndividualBehaviors2003, nowakFiveRulesEvolution2006}.

Here we examine another mechanism: conformist norm internalization, in which an agent conforms to the majority behavior of the group \cite{gintisHitchhikerGuideAltruism2003, lehmannCoevolutionCulturallyInherited2008}, even if this runs counter to its own (fitness) interest. Because those in cooperative groups thus conform to the cooperative norm, this conformity implies that cooperators are more likely to be in cooperative groups than non-cooperators. But this \emph{consequence} of norm internalization doesn't elucidate why a high population of norm internalization came about in the first place \cite{lehmannSocialIndividualLearning2008}. In this paper, we examine whether norm internalization can create conditions of higher cooperation \emph{without} making any assumptions about their population: that is, we allow for the possibility that they might die out.

More formally, we ask, assuming the presence of group selection and punishment,
\begin{itemize}
    \item[Q1] Can norm internalization and cooperation proliferate when both start out rare, and, if yes,
    \item[Q2] Does the presence of norm internalizers (NIs) increase cooperation levels over and above the effects of group selection and punishment alone?
\end{itemize}

\noindent
We split the question into two parts because we \emph{do not} take norm internalization as an exogenous given. Instead, we include it as a possible agent strategy, with the possibility wide open that the strategy might go extinct (this is true of any strategy in the strategy space). Thus, to have an effect that we care about, norm internalization must (1) not go extinct and (2) influence cooperation levels positively, which is by no means a given, since norm internalizers may internalize a defection norm (for more reasons see \cite{henrichWhyPeoplePunish2001, lehmannCulturalTransmissionCan2008}). This means that we are concerned both with the selective pressures in favor of norm internalization (its causes, Q1), and its consequences (Q2).

The reason norm internalization could have an impact above the already powerful effects of punishment \cite{boydPunishmentAllowsEvolution1992, 
henrichWhyPeoplePunish2001, boydEvolutionAltruisticPunishment2003, boydCoordinatedPunishmentDefectors2010} is that punishment, in our model, is imperfect: many defections go undetected and, furthermore, agents know how likely they are to be caught. In these cases, can norm internalization do what punishment on its own cannot: keep agents cooperating even when no one is looking?

Most previous work on the issue, in contrast to ours, (1) does not include group selection or punishment (and certainly not the imperfect punishment we use) \cite{lehmannCoevolutionCulturallyInherited2008}, (2) holds the norm internalization trait fixed, therefore not answering question (Q1) \cite{henrichWhyPeoplePunish2001, lehmannCulturalTransmissionCan2008}, (3) uses a non-conformist norm internalization mechanism, where the pressure to internalize a norm is instead specified exogenously \cite{gavriletsCollectiveActionEvolution2017, lozanoCooperationSocialNorm2020}, or (4) assumes some outside benefit to conformism \cite{gintisHitchhikerGuideAltruism2003}.

We hypothesized that norm internalization would increase between-group variation, thereby amplifying group selective forces \cite{henrichCulturalGroupSelection2004, boydOriginEvolutionCultures2005}. This would then favor cooperative groups, which we expected would contain more NIs --- positive feedback in favor of NIs. 

However, we found that while norm internalization polarized groups, high-cooperation groups did not necessarily have more NIs. Thus, the basic feedback hypothesis was too simple. 

Regarding Q1, we found that norm internalization didn't quite proliferate when rare, as only a minority of agents possessed the trait in all simulations we ran. Instead, the population of NIs averaged between a third and almost zero. Nevertheless, with regard to Q2 we found that NIs \emph{did} substantially increase cooperation levels, when both cooperation and norm internalization started out rare, over and above the effects of group selection and punishment. Importantly, this effect did not obtain when the NI population was the highest; rather, it was the most pronounced when NIs constituted on average about 10\% of the population.

How did such a small population of NIs increase cooperation levels so substantially? They sparked higher levels of cooperation in other agents, by playing the following roles:

\begin{itemize}[leftmargin=0.75in]
    \item[\textit{Polarizer}] A high prevalence of norm internalizers (NIs) in a group tended to lead either to extreme cooperation or extreme defection among its members, amplifying intergroup differences in cooperation levels.
    \item[\textit{Catalyst}] NIs were especially effective at precipitating cooperation (among all agent types) when global cooperation levels started very low.
    \item[\textit{Stabilizer}] The presence of NIs in groups tended to prolong bouts of high total cooperation.
\end{itemize}

We derived our results from agent-based simulations where agents (1) play a public goods game in groups and (2) evolve over generations, with cooperation always starting rare. We addressed Q1 by starting with a low NI population and observing the results --- this enabled us to see whether NIs could survive without going extinct, and what population level they attained. As for Q2, we compared two conditions: versions of the simulation that included NIs in the strategy space (with nothing preventing the small starting population from going extinct) with ones that did not (starting instead with a small population of unconditional cooperators). This helped us to parse whether norm internalization was able to increase cooperation levels beyond what group selection and punishment could do alone. 

We performed these tests with two very different agent-based models (fully specified in \ref{appendix:model-fuller-specification} with parameters in \ref{appendix:results-parameters}), choosing this approach to show that our results, which turned out to be quite similar for the two models, do not depend on specific implementation details or assumptions about, for instance, group conflict rates. 

Section~\ref{section:background} will provide further background on norm internalization, group selection, and punishment, the three interacting `prosocial forces' in our model. Then, Section~\ref{section:models} describes the models, and Section~\ref{section:results} presents the results, whose implications are discussed in Section~\ref{section:discussion} and Section~\ref{section:conclusion}.

\section{Background} \label{section:background}

Most of the literature on norm internalization focuses on its consequences for the evolution of cooperation, rather than on its causes. 

Here, we are concerned with both the causes and consequences of norm internalization, with Q1 concerned with the former and Q2, the latter.

In this section, we first define norm internalization and review the literature on its causes (the selective pressures favoring it) and consequences. We also examine how the two other forces in favor of cooperation, group selection and punishment, interact with norm internalization.

\subsection{Defining norm internalization} \label{section:background-defining}

We define norm internalization as the tendency to follow the majority behavior of the community, even at the expense of one's own fitness. This is closely linked to the notion of conscience, which is an internal enforcer of norm-following that is powered by emotions such as guilt and shame \cite{tangneySelfconsciousEmotionsShame2005, frithWhatUseConsciousness2016}. For instance, Boehm describes conscience as the ``internalization of values'' \cite{boehmMoralOriginsEvolution2012}, and Churchland acknowledges that conscience involves ``feelings that urge us in a general direction, and judgment that shapes the urge into a specific action'' \cite{churchlandConscienceOriginsMoral2020}. However, because `conscience' is bound up with moral emotions that are not explicitly represented in our models, we will stick to the term `norm internalization'.

Empirically, humans internalize norms from those around them \cite{parsonsSociologicalTheoryModern1967, grusecImpactParentalDiscipline1994}, and follow norms even when not being observed \cite{fischbacherLiesDisguiseExperimental2013}. Children, furthermore, are `promiscuous normativists', ascribing a normative character indiscriminately to observed actions, even going so far as to enforce behavior in accordance with those actions \cite{schmidtYoungChildrenSee2016}. In what follows, we aim to elucidate the evolutionary origins of these empirically observed traits.

\subsection{Possible causes of norm internalization} \label{section:background-possible-causes}

Norm internalization as described in Section~\ref{section:background-defining} is a type of \emph{conformist transmission} (defined as copying prevalent behaviors in the group \cite{henrichCulturalGroupSelection2004}), which Henrich and Boyd showed to be beneficial in a noisy environment, as it allows individuals to effectively base their decisions on a large number of samples from the environment -- that is, the samples taken by everybody else, and not just their own. Because a larger set of samples better approximates the ground truth, conformist behaviors can lead to better-adapted outcomes \cite{henrichEvolutionConformistTransmission1998}. Coordination games also provide selection pressure in favor of conformist transmission \cite{youngSocialNormsEconomic1998, mcelreathSharedNormsEvolution2003}.

This beneficial aspect of conformism has led to the exaptation hypothesis, which states that humans developed a propensity to copy others because it is usually beneficial, but individually-harmful altruistic behaviors are also copied `by mistake' (in evolutionary terms) \cite{henrichWhyPeoplePunish2001, gintisHitchhikerGuideAltruism2003}. Still, copying behaviors would result in a net benefit for fitness, given the quantity of cultural behaviors that people simply could not learn on their own \cite{henrichEvolutionCulturalEvolution2003}.

The exaptation hypothesis has a shortcoming. Many altruistic actions are manifestly costly, in terms of time (participating in group hunts), forgone benefits (not stealing), or a risk to one's life (going to war). For this reason, the assumption that an agent would learn to copy behaviors indiscriminately as a noise-reduction mechanism seems too strong: more plausibly, conformist transmission helps the agent decide among multiple options about which some information is available. Thus, in this paper, we don't assume any external benefits to the norm internalization trait (such as the ability to learn from others in a noisy environment).

Another proposed cause of the tendency to abide by the cooperation norm, even when there would be no future consequences to defection, is \emph{risk-management}. In a world full of agents using the tit-for-tat strategy \cite{axelrodEmergenceCooperationEgoists1981}, it is much more costly to defect when one should have cooperated (since it leads to an endless chain of future defections) than it is to cooperate when one could have defected (since this simply involves a one-time payment of the cooperation cost, \cite{deltonEvolutionDirectReciprocity2011}). This is an elegant explanation of norm-following under certain conditions, but it is not quite sufficient for the scenario we describe, for two reasons. First, it assumes a world of tit-for-taters, when there is no particular reason for doing so, since tit-for-tat is not an evolutionarily stable strategy (no pure strategies are stable in the iterated prisoner's dilemma \cite{lorberbaumNoStrategyEvolutionarily1994}). Second, in n-person interactions, which is what we examine here with a public goods game, direct reciprocity in the style of tit-for-tat quickly collapses \cite{boydEvolutionReciprocitySizable1988}.

\subsection{Group selection: Consequences of norm internalization and conformist transmission} \label{section:introduction-group-selection}

Both norm internalization and conformist transmission have an important consequence: they can amplify group selection. Group selection is vital in sustaining cooperation \cite{wilsonMultilevelSelectionTheory2008}, because it can offset the individual's loss from cooperation with benefits accrued by the group. The strength of group selection depends on the variance among groups with respect to the trait in question \cite{henrichCulturalGroupSelection2004}, and as a result, it is particularly effective when groups are small, so that they deviate more from the average by the law of large numbers; when migration rates are low, so that differences among groups are maintained \cite{maynardsmithGroupSelectionKin1964}; and, importantly, when conformist transmission is strong (imagine two groups respectively with 40/60 and 60/40 cooperation/defection ratios -- if conformist transmission occurs, these groups will respectively move toward full defection and full cooperation as agents copy the majority behavior, increasing inter-group variation \cite{henrichEvolutionConformistTransmission1998}). Thus, conformist transmission, with this polarizing tendency, could boost the evolution of cooperation \cite{fehrStrongReciprocityHuman2002, henrichCulturalGroupSelection2004, boydOriginEvolutionCultures2005}, though with some caveats \cite{lehmannCulturalTransmissionCan2008}.

\subsection{Co-evolutionary hypothesis as a cause of norm internalization} \label{section:background-coevolutionary-hypothesis}

This polarizing consequence of norm internalization leads us to the hypothesis that norm internalization becomes entrenched by amplifying group selection and co-evolving with cooperation. Several studies have explored this hypothesis, with mixed results:

Gavrilets and Richerson found that NIs evolved when there is strong pressure to punish free-riders in part because, by cooperating, they allow the group to save on the cost of punishment \cite{gavriletsCollectiveActionEvolution2017}. However, their version of norm internalization was not dependent on the frequency of the trait --- instead, the social pressure was exogenously determined by a model parameter. Our paper differs in two important respects: first, their punishment mechanism (which was central to their result) was perfect, in that all defectors could be punished. In our study, we are interested in the possibility that norm internalizers might fill the gap of imperfect punishment, where not all agents are caught, and thus we focus on the internalization of the \emph{cooperation} norm --- which could potentially keep agents cooperating even when no potential punishment is observing them. Further, while Gavrilets et al (and other more recent papers, such as \cite{lozanoCooperationSocialNorm2020}) study a version of norm internalization whose strength is specified by an exogenous parameter, our experiment examines the case where the social pressure to follow the norm depends on the percentage of group members that follow it --- leading to very different dynamics.

Lehmann and Feldman found that generally, conformist transmission of helping behaviors (which can be interpreted as norm internalization) does not promote culturally transmitted cooperation \cite{lehmannCoevolutionCulturallyInherited2008}. One reason for that could have been the model's assumption that groups completely reshuffle every generation, which severely hampers the development of inter-group differences. In contrast, in this paper, we relax this assumption, and also include punishment by indirect reciprocity.

\subsection{Punishment by indirect reciprocity}

With group selection, norm internalization has an amplification effect, and with punishment, it has a complementary effect. Norm internalization is cheap when cooperation is rare, because (costly) cooperative norms are only internalized when a majority of agents cooperate. In comparison, punishment is expensive when cooperation is rare: having many defectors means that the population must expend more resources on meting out punishment \cite{boydEvolutionAltruisticPunishment2003} (see \cite{boydCoordinatedPunishmentDefectors2010} for a possible solution). The inverse is true, too: norm internalization is expensive when cooperation is common (as it leads to the internalization of a costly cooperation norm), while punishment is cheap (few defectors to punish). Furthermore, norm internalization can get agents to cooperate where the threat of punishment cannot: that is, when an action is unlikely to be observed, and therefore unlikely to be punished. Taken together, these contrasting characteristics suggest that norm internalization and punishment would effectively complement each other.

In our model, the punishment mechanism is an imperfect variant of indirect reciprocity, namely, the withholding of benefits from those caught defecting --- imperfect because not all defections are observed (cf. \cite{hirshleiferCooperationRepeatedPrisoners1989}). This form of punishment is not costly for the punisher, and it creates an incentive for egotistic agents to cooperate at least some of the time. This practice is evolutionarily stable under certain conditions \cite{panchanathanTaleTwoDefectors2003, ohtsukiLeadingEightSocial2006, odouardTitTattlingCooperation2023}, is observed in many societies \cite{henrichFairnessPunishmentBehavioral2014, bhuiHowExploitationLaunched2019}, and is reproduced in experiments \cite{wedekindCooperationImageScoring2000}. For all these reasons, we \emph{fix} the punishment mechanism, focusing not on the co-evolution of punishment and norm internalization, but rather on whether norm internalization can do its job, given the existence of an imperfect punishment mechanism.

\subsection{Our niche}
In brief, then, in contrast to Lehmann et al.\ and Gavrilets and Richerson, our study focused on the interaction between frequency-based NIs, group selection, and `imperfect' punishment. Further, no external benefits to agents are assumed (contra \cite{gintisHitchhikerGuideAltruism2003}), and no particular strategy is assumed (contra \cite{deltonEvolutionDirectReciprocity2011}). Could this combination of factors lead both norm internalization and cooperation to proliferate when rare?

\section{The models} \label{section:models}

To address our questions, we designed agent-based models with free-floating populations of various agent strategies, in which both norm internalization and cooperation started rare. No version of the model had a fixed distribution of strategies --- their relative frequencies were always free to evolve. Our objective was to measure (Q1) the equilibrium population of norm internalizers (NIs) --- did they go extinct? sweep to fixation? and (Q2) the differences in cooperation levels when NIs were and were not present in the strategy space (allowing us to measure the effect of NIs over and above group selection and punishment).

We tested two models\footnote[2]{The code for the model, along with the data it generated, can be found at \texttt{\url{https://github.com/victorvikram/norm-internalization-and-coop}}} because agent-based simulations can yield very different results with only minor differences in implementation (see \cite{nowakEvolutionaryGamesSpatial1992, hubermanEvolutionaryGamesComputer1993,galanAppearancesCanBe2005}). The first, `abstract model' bears many similarities to previous work \cite{boydEvolutionAltruisticPunishment2003, nowakFiveRulesEvolution2006}: co-location of groups, fixed group sizes, fixed numbers of groups, etc. This makes it easier to both compare it with other models. The second, `naturalistic model' is more complex, locating groups in space, inspired by \cite{grimmPatternorientedModelingAgentbased2005}. We give an overview of the model in this section; \ref{appendix:model-fuller-specification} contains a full specification.

\subsection{Basics} \label{section:model-basics}

Both models follow the same series of steps:
\begin{itemize}
\item \label{item:decision}\emph{Decision} --- Agents decide, based on their strategy, whether to cooperate or defect.
\item \label{item:distribution}\emph{Distribution} --- Public benefits are distributed to agents that were not caught defecting (punishment occurs in this step).
\item \label{item:igd}\emph{Inter-group dynamics} --- Groups compete, either indirectly for resources or directly by conflict. Migration occurs.
\item \label{item:selection}\emph{Selection} --- Individuals survive and reproduce according to fitness.
\end{itemize}

\subsubsection{Decision} \label{section:model-basics-decision}

Agents, divided into groups, decide whether to cooperate or defect in a public goods game \cite{boydEvolutionAltruisticPunishment2003}. To cooperate means to pay a cost, $c$, to produce a benefit, $b > c$, that will be shared among the group \cite{talhelmLargescalePsychologicalDifferences2014}. Agents also have a probability of being observed, which is sampled uniformly at random, for each agent, every round. This probability can influence agent decisions.

Agents decide whether to cooperate using a strategy parameterized by two variables: (1) the propensity to cooperate, $\pi$, and (2) the learning style, which defines how $\pi$ changes. The different values of $\pi$ produce a continuous space that includes unconditional defection, expected value (EV) maximization, and unconditional cooperation. Norm internalization is encoded in the learning style.

More specifically, with $p_{\text{obs}}$ being the probability of being observed and $\overline{b}$ the average group benefit distribution (see Section~\ref{section:model-basics-distribution}) in the previous round, an agent cooperates when

\begin{equation} \label{eqn:cooperation-inequality-spatial}
p_{\text{obs}}\overline{b} \geq (1 - \pi)c,
\end{equation}

\noindent
unless they err, with probability $\epsilon$. Thus, $\pi = 0$ corresponds to an EV-maximizer, as the resulting equation compares the expected cost of defecting ($p_{\text{obs}}\overline{b}$) to the expected cost of cooperating  ($c$) and picks the lower-cost option. By contrast, $\pi = 1$ corresponds to unconditional cooperation (since $p_{\text{obs}}\overline{b} \geq 0$ no matter what), and $\pi \ll 0$ corresponds to unconditional defection. The learning style defines how $\pi$ changes:

\begin{itemize}
    \item \emph{Norm internalization} - Gradually approach a $\pi$ value of $1$ (cooperation regardless of who is watching) as long as a majority of other agents cooperate; otherwise, approach $0$ (EV-maximization).
    \item \emph{Selfish} - Move in the direction that yields the best individual payoff.
    \item \emph{Static} - Don't learn. 
\end{itemize}

\noindent
We included the selfish learners to strengthen our results, as individual learning likely hampers the evolution of cooperation \cite{lehmannSocialIndividualLearning2008}. Further, it ensures NIs are not be the only agents with a dynamic  $\pi$-value, providing them with adequate competition from a selfish agent that also had a dynamic $\pi$-value. Of course, many selfish learning rules are possible, we chose this one for simplicity --- our focus is to compare the presence and absence of NIs in the strategy space, not the effects of different types of selfish strategies.

\subsubsection{Distribution}\label{section:model-basics-distribution}

After agents make their choices, they are observed with probability $p_{\text{obs}}$. The total contributions of cooperators are distributed to group members who were not observed defecting, each receiving a share $\overline{b}$. Thus, agents for whom there was either (1) a high $p_\text{obs}$ or (2) a high $\overline{b}$ have a stronger incentive to cooperate. The distribution of the benefits results in the payoffs shown in Table~\ref{table:abstract-payoff} (see also Fig.~\ref{fig:model-summary}.a2,n3):

\begin{table}
\begin{center}
\caption{\textbf{Payoffs in the model}. The $f$ is the baseline fitness level that all agents receive; agents pay $c$ (the cost of cooperation) if they choose to cooperate. The group distribution, $\overline{b}$, depends on how many agents in the group decided to cooperate: it is calculated as the total public benefit divided by the number of agents who received a share
}
 \begin{tabular}{c c c}
        & Defect & Cooperate  \\
        \hline
        Observed & $f$ & $f - c + \overline{b}$ \\[0.6em]
        \hline
        Unobserved & $f + \overline{b}$ & $f - c + \overline{b}$ \\
    \end{tabular} \label{table:abstract-payoff}
\end{center}

\end{table}

\subsubsection{Inter-group dynamics and selection}\label{section:model-basics-igd-selection}

Inter-group conflict is implemented quite differently in the two models, as described in detail in the following Sections~\ref{section:abstract-model} and~\ref{section:naturalistic-model}. As for migration, we test a variety of migration rates, but in both models focus most of our analysis on parameters that yield approximately a 50\% migration rate -- that is, agents have a 50\% probability of dying in a group that they were not born in, motivated by the high migration probabilities observed in hunter-gatherer societies \cite{hillCoresidencePatternsHuntergatherer2011}.
\begin{table}
    \caption{\textbf{Key parameters in both models.} Strictly speaking, in the naturalistic model, $n$ and $g$ are the starting number of individuals in a group, and the starting number of groups. Parameters in brackets were only tested in the naturalistic model}
\begin{center}
   \begin{tabular}{c p{6cm} c | c}
        Variable & Description & Abstract & Naturalistic  \\
        \hline
        $n$ & number of individuals per group & \texttt{35} & \texttt{20} \\
        $g$ & number of groups & \texttt{60} & \texttt{10}  \\
        $p_{\text{con}}$ & probability of conflict & \texttt{1/13} & \texttt{N/A} \\
        $s$ & side length of square grid & \texttt{N/A} & \texttt{10} \\
        $c_{\text{dist}}$ & cost of foraging on an adjacent square & \texttt{N/A} &\texttt{5} \\
        $b/c$ & benefit to cost ratio & \multicolumn{2}{c}{\texttt{3, 3.25, 3.5, 3.75, 4.0, 4.25}} \\
        $p_{\text{mig}}$ & probability of migration & \multicolumn{2}{c}{\texttt{0, 0.1, ... 0.6, [0.7, 0.8]}} \\
    \end{tabular} 

    \label{table:important-parameters}
\end{center}
\end{table}
\subsubsection{Parameters}

We ran both models on the lower end of the range of values in which cooperation did not go extinct, which was $b/c \in [2.75, 4.25]$, with increments of 0.25 (benefits lower than this range never resulted in cooperative worlds). We tested two conditions:

\begin{itemize}
\item \textit{Norm internalization} - 2\% of agents started with the norm internalization trait.
\item \textit{No norm internalization} - No NIs were present in the strategy space, so we replaced the 2\% starting population of NIs with unconditional cooperators.
\end{itemize}

\noindent
For both conditions, the remainder of agents were split between selfish and static learners, with each with a (unique) propensity $\pi$ uniformly sampled between $-1$ and $1$. We did this to make as few assumptions about the initial population as possible. That said, agents with $\pi$-values far from zero quickly died off, leaving behind, at the start of the run, agents who were effectively EV-maximizers. 

As for why we included unconditional cooperators in the non-NI model, we needed a cooperative-type of agent for forces like group selection and punishment to favor, to fill the role of the NIs in the NI-condition. The obvious choice to fill this role is the unconditional cooperator.

 The key parameters appear in Table~\ref{table:important-parameters}. For a full list, along with run counts, see \ref{appendix:results-parameters}.

\subsection{Abstract model} 
\label{section:abstract-model}

In the abstract model, the decision and distribution steps are exactly as described in Section~\ref{section:model-basics}.

\begin{center}
    \begin{figure}[htbp]
    \centering
    \includegraphics[width=14cm]{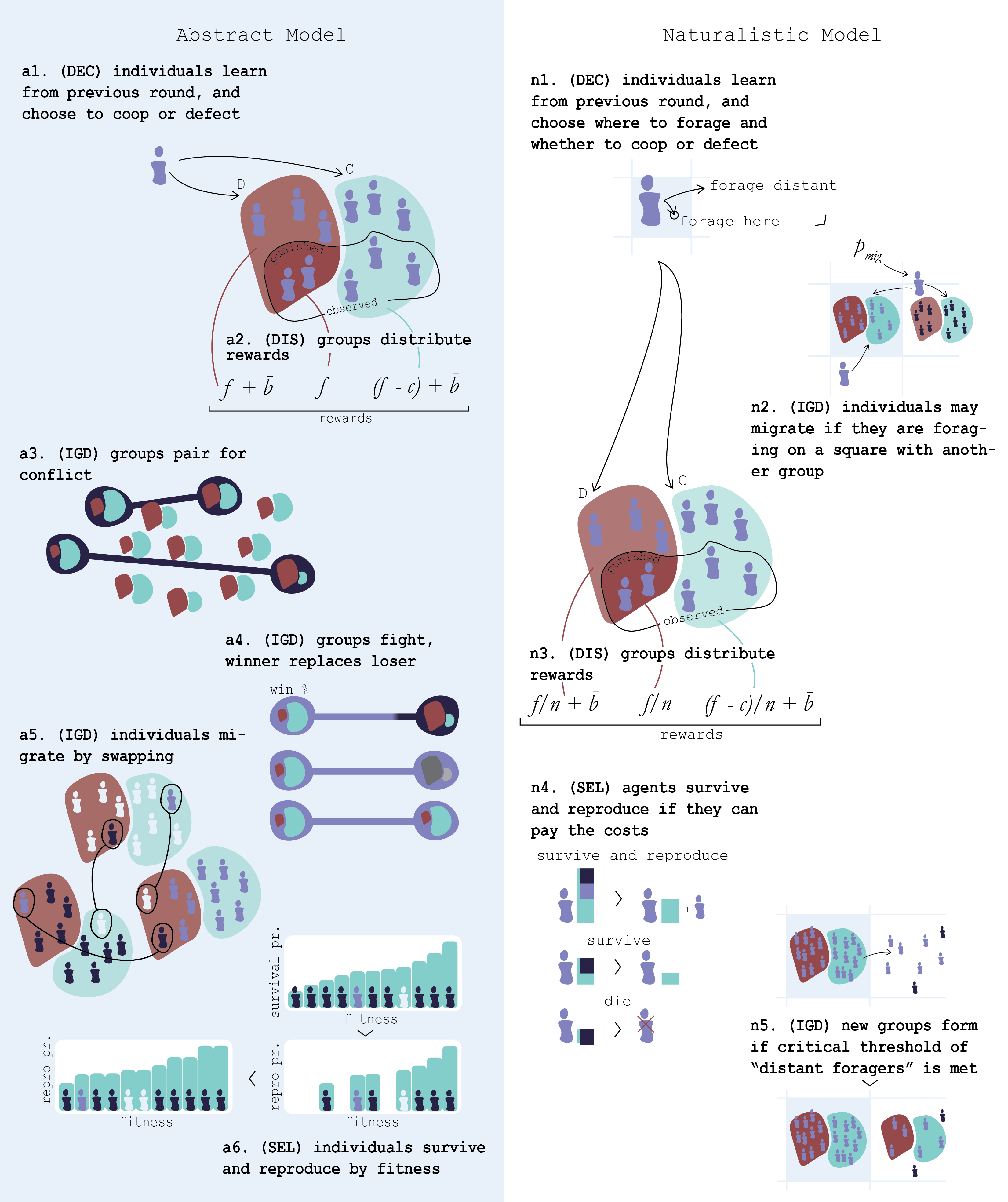}
    \caption{\textbf{The steps taken in each iteration of the model.} Described fully in the text. In the abstract model, decision (\texttt{DEC}), distribution (\texttt{DIS}), inter-group dynamics (\texttt{IGD}), and selection (\texttt{SEL}) occur in sequence. By contrast, in the naturalistic model these steps are interspersed with each other}
    \label{fig:model-summary}
\end{figure}
\end{center}

Regarding intergroup dynamics, groups pair up (Fig.~\ref{fig:model-summary}.a3), and engage in conflict with probability $p_\text{con}$. Groups with higher average fitness are more likely to win (Fig.~\ref{fig:model-summary}.a4). Following Bowles, we set $p_\text{con}$ to approximately one conflict every four agent lifetimes \cite{bowlesIndividualInteractionsGroup2001}.

Then, in the migration step, individual agents pair up and switch groups with probability $p_\text{mig}$ (Fig.~\ref{fig:model-summary}.a5). For more details, see \ref{appendix:abstract-intergroup}.

As for the (individual) selection phase, a fixed number of agents survive and reproduce every round, chosen probabilistically according to their fitness (Fig.~\ref{fig:model-summary}.a6). For more details, see \ref{appendix:abstract-selection}.

\subsection{Naturalistic model}\label{section:naturalistic-model}

In contrast to the abstract model, the naturalistic model locates groups on a spatial grid. The spatial aspect causes groups to compete by encroaching on each other's foraging grounds, obviating the need for a direct conflict component (Fig.~\ref{fig:model-summary}).

\subsubsection{Decision}\label{section:naturalistic-decision}

In addition to the decision on whether to cooperate (Section~\ref{section:model-basics-decision}), agents also decide where to forage (Fig.~\ref{fig:model-summary}.n1), knowing that the more foragers there are on a square, the lower their payoff (due to competition; Section~\ref{section:naturalistic-distribution}). An agent calculates the expected payoff of foraging on their home square compared to that of foraging on a randomly chosen adjacent square and chooses where to go with probabilities proportional to these payoffs (there is an additional ``cost of foraging on distant lands,'' $c_\text{dist}$ so even if one's home square is more crowded, it may still pay to stay put).

\subsubsection{Distribution} \label{section:naturalistic-distribution}

The only addition to what is described above is that an agent's payoff is scaled by the number of agents foraging on the grid square. That is, both the private and public benefit generated by an agent are divided by $n_\text{here}$, the number of foragers on that agent's square.

\subsubsection{Intergroup dynamics}

In this model, groups compete for resources without direct conflict. Groups grow when their members have high fitness (Section~\ref{section:naturalistic-reproduction}), leading to crowding, so there is a mechanism by which a new group can bud off an old group (Fig.~\ref{fig:model-summary}.n6). Recall that agents need not necessarily forage on their group's home square (Section~\ref{section:naturalistic-decision}). Generally, when they forage on another square, they remain tied to their group. However, if at least $n$ agents from a group are foraging on a square other than their home, and they make up the plurality of agents on that square, then they start a new group on that square if none is already there. In this manner, populous groups can spread across the landscape.

Because the benefit derived from foraging is inversely proportional to the number of agents present, these new groups begin to compete for the resources of pre-existing groups. Groups that make the most of available resources (the cooperative ones) are therefore expected to out-compete the others.

Finally, if an agent is foraging away from their group's home square, and there happens to be another group on their current square, they may migrate to that group with probability $p_\text{mig}$ (Fig.~\ref{fig:model-summary}.n2).

\subsubsection{Selection}\label{section:naturalistic-reproduction}

In this model, there are costs of surviving and costs of reproducing. An agent's ability to pay the cost of surviving (resp. reproducing) determines whether they survive (resp. reproduce), as shown in Fig.~\ref{fig:model-summary}.n5. For a group to grow, it must have a significant proportion of agents with high enough fitness to pay both costs. Thus, high group-average fitness leads to growth, in contrast to the abstract model, where group size is constant and high group-average fitness leads to an advantage in direct conflict. For details, see \ref{appendix:naturalistic-selection}.

\section{Results} \label{section:results}

Despite all the differences in implementation between the abstract and naturalistic models, both models exhibited remarkably similar results. In what follows, we will flag qualitatively different results with a \emph{difference} annotation. But, in both models,

\begin{enumerate}
\item norm internalizers (NIs) made up a small minority of the population. In fact, for the parameter settings under which they made the biggest difference in cooperation levels, they constituted an average of 10\% of the population.

\begin{figure}[htbp]
    \centering
    \centerline{\includegraphics[width=18cm]{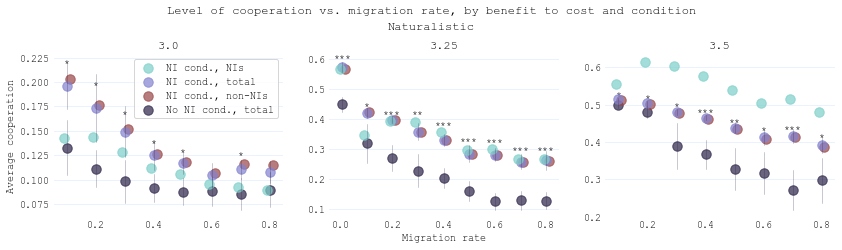}}
    \centerline{\includegraphics[width=18cm]{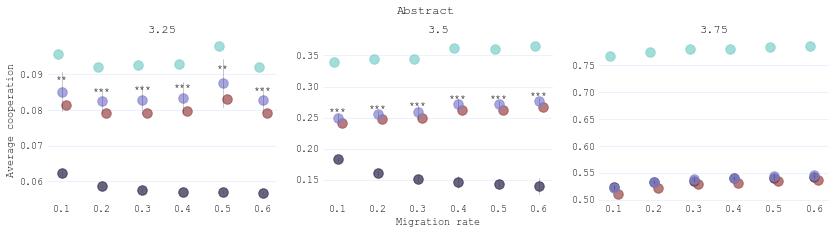}}
    \caption{\textbf{Levels of cooperation for various model conditions.} Shown are the mid-range benefit-to-cost ratios (e.g., \texttt{3.0}, \texttt{3.25}, ...) for which the norm internalizer condition (NI cond.), that is, the condition in which norm internalizers (NIs) were in the strategy space, had significantly higher level of cooperation. For benefit-to-cost ratios above or below the ones shown, the NI condition did not have significantly different cooperation levels. \newline
    Some observations to note are \textbf{(1)} Whenever the NI condition had significantly different levels of cooperation, it was always higher. \textbf{(2)} NIs did not necessarily cooperate more than non-NIs, in fact, in the naturalistic model, NIs made the biggest difference in cooperation levels when they cooperated about equally with everyone else. \textbf{(3)} As the benefit-to-cost ratio rises, the level of NI-cooperation rises comparatively to non-NIs. \textbf{(4)} The fall-off in cooperation with increasing migration rates is less steep in the NI condition -- in fact, in the abstract model, there is no fall-off at all. \newline
    Full parameter specifications are in \ref{appendix:results-parameters}. In all figures, except when axes indicate otherwise, we use the \texttt{default} parameter set, $b/c = \texttt{3.5}$, $p_\text{mig} = \texttt{0.2}$ for the abstract model, and $b/c = \texttt{3.25}$, $p_\text{mig} = \texttt{0.5}$ for the naturalistic model (these two migration rates end up being equivalent at about a 50\% lifetime migration rate); see \ref{appendix:migration-remark} for more justification. \newline
    The error bars are 95\% confidence intervals ($t$-test), shown only for the total cooperation in each of the two conditions. (*), (**), and (***) respectively indicate the difference between the NI and non-NI condition are different with $p < 0.01$, $p < 0.001$, and $p < 0.0001$ (two-tailed $t$-test)}
    \label{fig:summary}
\end{figure}

\begin{figure}
    \centering
    \centering
    \begin{subfigure}{.5\textwidth}
        \centering
        \includegraphics[width=7cm]{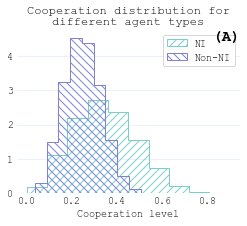}
    \end{subfigure}%
    \begin{subfigure}{.5\textwidth}
      \centering
      \includegraphics[width=6.5cm]{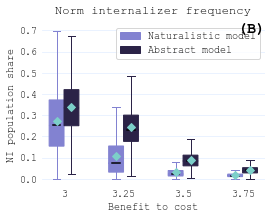}
    \end{subfigure}
    \caption{\textbf{Norm internalizer cooperation and population.} \textbf{(A)} Histogram showing the density of cooperation percentages of norm internalizers (NIs) vs non-NIs. NI cooperation tends to be more dispersed in intermediate benefit-to-cost regimes. See \ref{appendix:counterpart-figures} for the equivalent figure in the naturalistic model, exhibiting a similar pattern. \textbf{(B)} The average population share of NI across rounds, in the naturalistic and abstract models. For reference, NIs made the biggest difference in cooperation at \texttt{3.25} in the naturalistic model and \texttt{3.5} in the abstract model. The diamonds are the means. The 95\% confidence intervals for the median are represented by notches on the box plot, which are so small as to be invisible. Parameters used are the \texttt{default} set, except for where $b/c$ is otherwise specified}
    \label{fig:norm-internalizer-quick-stats}
\end{figure}

\item NIs had bouts of high cooperation that were significantly \emph{shorter} than those of the other agents (though they did go higher). 
\item The presence of NIs did not lead to higher levels of cooperation in groups in which they were more common
\end{enumerate}

\noindent        
Yet, we found that for mid-range benefit-to-cost ratios, the norm internalizer condition had

\noindent
\begin{enumerate}
\item \emph{higher} levels of cooperation
\item  \emph{longer} peaks of above-average cooperation.
\item fewer long bouts of below-average cooperation (though this held only in the naturalistic model).
\end{enumerate}

To shed light on this counterintuitive result, we show that norm internalizers

\begin{enumerate}
\item \emph{polarize} groups either to extreme cooperation or extreme defection, perhaps enhancing group selective forces that favor cooperative groups;
\item  tend to be the ones to \emph{catalyze} bouts of above-average cooperation in all agents -- not just other NIs -- when cooperation is especially low (this result was stronger in the naturalistic model);
\item  help to \emph{stabilize} high levels of cooperation, keeping peaks high for longer.
\end{enumerate}

\subsection{Average cooperation in each condition}

The presence of norm internalizers (NIs) in the strategy space (the NI-condition) either increased mean levels of cooperation or made no difference. In Fig.~\ref{fig:summary}, we show the range of benefit-to-cost (b:c) ratios for which the presence of NIs did make a difference: the mid-range. That is, there is a low range of b:c ratios for which surplus cooperation is essentially zero (surplus cooperation is the amount of cooperation above the 5\% error rate), and a high range for which relatively high cooperation levels emerge, regardless of whether NIs are present. But for ratios in between those extremes --- from at least \texttt{3 - 3.5} in the naturalistic model and \texttt{3.25 - 3.5} in the abstract model --- NIs made a significant difference. In the majority of these mid-range conditions, their presence at least doubled the surplus cooperation.

Despite their tendency to boost cooperation, NIs did not necessarily cooperate more than other agent types. Even if they did, most of the boost in average cooperation was due to NIs catalyzing high cooperation among \emph{other} agent types: in Fig.~\ref{fig:summary}, the majority of the difference in total cooperation in the NI condition (medium dots) and the non-NI condition (dark dots) is accounted for by the rise in cooperation of non-NIs. 

\begin{center}
\begin{figure}[htbp]
  \centering
  \begin{minipage}{\linewidth}
    \begin{subfigure}[r]{0.5\linewidth}
        \raggedleft
      \includegraphics[width=3.25cm]{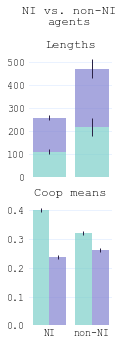}
      \label{fig:subfigure1_label}
    \end{subfigure}%
    \begin{subfigure}[l]{0.5\linewidth}
      \raggedright
      \includegraphics[width=3.25cm]{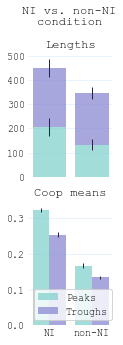}
      \label{fig:subfigure2_label}
    \end{subfigure}
    \caption{\textbf{Spike-trough comparison for the naturalistic model.} Comparisons of the \emph{lengths} and \emph{cooperation means} of peaks and troughs. Peaks are defined as intervals of at least ten rounds in which cooperation is above average; troughs are the complement of that. At \textbf{left}, we compare between cooperation peaks of \emph{norm internalizers} and cooperation peaks of other agent types, all \emph{within the norm internalizer condition}. At \textbf{right}, we compare between the NI-condition (NIs are in the strategy space) and the non-NI condition. Though the peaks of norm internalizers are shorter (top left), the peaks in the norm internalizer condition are \emph{longer} (top right). And though the trough cooperation of norm internalizers is slightly lower (bottom left), the trough cooperation levels in the norm internalizer condition are higher (bottom right). 
    Error bars are 99\% confidence intervals ($z$-test), both cooperation means and lengths are significantly different at that level. Parameters are the \texttt{default} set.}
    \label{fig:spike-summary}
  \end{minipage}
\end{figure}
\end{center}

In fact, in the naturalistic model, the NI-condition boosted cooperation even when NIs tended to cooperate \emph{less} than other agent types (with $b/c = 3$), and the NI condition made the most difference when NIs cooperated about equally ($b/c = 3.5$). In the abstract model, however, NIs always cooperated more on average than their counterparts (one notable \emph{difference} between the models). That said, both models shared the characteristic that NIs had a higher variance in their cooperation levels, which can be seen in Fig.~\ref{fig:norm-internalizer-quick-stats}a.

The drop-off in the NI's ability to catalyze cooperation for larger b:c ratios likely has to do with with the drop in their population as b:c rises. This is because, as b:c rises, the extent to which NIs cooperate disproportionately also grows (see the light dots in Fig.~\ref{fig:summary}). This causes their population to take a hit (see Fig.~\ref{fig:norm-internalizer-quick-stats}b). With that said, at no b:c did NIs make up a majority of the population (averaged across rounds). In fact, their population was only around 10\% for the ratios at which they were \emph{most} effective at bringing about higher levels of cooperation (\texttt{3.25} in the naturalistic model and \texttt{3.5} in the abstract model).

\subsection{Shape of cooperation in each condition}

The average only tells part of the story, however, especially because in our simulations, levels of cooperation tended to rise and fall cyclically. Two factors, therefore, could have increased the average: either the NI condition had relatively longer peaks (compared to its troughs) or it had higher average peak (and/or trough) cooperation levels. 

As can be seen in Fig.~\ref{fig:spike-summary} (on the right), the NI-condition exhibited both properties. That is, the peaks were longer and they had higher levels of cooperation. Further, a larger portion of the round was spent in the peak, but this difference was not significant. We define a peak as an interval in which cooperation is above average for at least ten rounds, and a trough as the complement of that.

This result stands in contrast to the cooperation peaks of the NIs themselves (at left). NI cooperation peaks, while occupying about the same proportion of the round, were much shorter. And while NIs cooperated more on their peaks, they cooperated less in their troughs, which meant that NIs cooperated barely (and certainly not significantly) more than non-NIs overall (\emph{difference}: in the abstract model, NIs did cooperate more. See Fig.~\ref{fig:abstract-spike-summary} in \ref{appendix:counterpart-figures} for the equivalent figure in the abstract model).

\subsection{The roles of norm internalizers}

We are left, then, with a puzzle. NIs themselves do not exhibit longer peaks of cooperation, nor do they necessarily cooperate more on average than other types of agents. The presence of NIs in the population must, therefore, contribute to longer peaks and higher cooperation levels indirectly.

Indeed, as the rest of this section shows, NIs play three vital roles in facilitating cooperation.

\subsubsection{Polarize} \label{section:results-polarize}

The central role of the norm internalizer is that of \emph{polarization}. High levels of NIs in a group caused cooperation to cluster at the extremes: either very low or very high. This makes sense: when cooperation is high, NIs internalize the tendency to cooperate, further increasing cooperation levels. The opposite is true at low levels of cooperation. We address in the discussion the intimate connection that polarization has with catalysis and stabilization (Sections~\ref{section:discussion-catalyze} and~\ref{section:discussion-stabilize}).

\begin{center}
    \begin{figure}[htbp]
        \centering
        \centerline{\includegraphics[width=17cm]{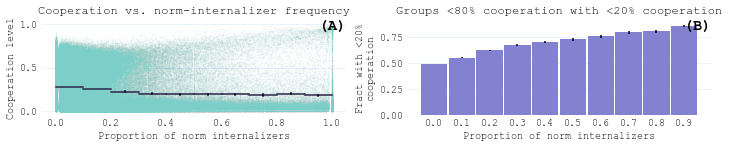}}
    \caption{\textbf{Polarization due to norm internalizers.} \textbf{(A)} Scatterplot of cooperation levels of individual groups at difference norm internalizer frequencies in the naturalistic model. The bars are the average cooperation levels for each window. The frequency of norm internalizers does not increase mean cooperation (in fact, there is a mild decrease) but it does polarize cooperation. \newline
    \textbf{(B)} The percentage of groups with cooperation levels below 80\% that \emph{also} have cooperation levels under 20\%. \newline
    Error bars are 99.9\% confidence intervals ($z$-test), parameters are the \texttt{default} set. For parallels with findings from the abstract model, see~\ref{appendix:counterpart-figures}}
    \label{fig:polarize}
\end{figure}
\end{center}

Polarization provides the first hint as to why NIs themselves do not have higher average cooperation levels: norm internalizers exhibit more extreme, not necessarily more cooperative behavior. Indeed, Fig.~\ref{fig:polarize} shows that higher populations of norm internalizers in a group do not increase the mean level of cooperation. This effect --- that high norm internalizer populations do not imply higher cooperation --- is the converse of the phenomenon observed earlier --- that higher cooperation does not imply higher norm internalizer populations (Fig.~\ref{fig:norm-internalizer-quick-stats}b).
\subsubsection{Catalyze}
The second role of the norm internalizer is its ability to catalyze cooperation, allowing it to bootstrap itself from extremely low starting points. This role was much more pronounced in the naturalistic than the abstract model (an important \emph{difference} between the models). We will propose reasons for why this might be so in Section~\ref{section:discussion-catalyze}; for now, we focus on the evidence from the naturalistic model. 

\begin{center}
    \begin{figure}[htbp]
    \centering
    \includegraphics[width=13cm]{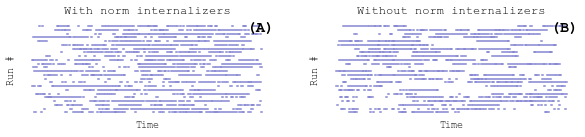}
    \includegraphics[width=13cm]{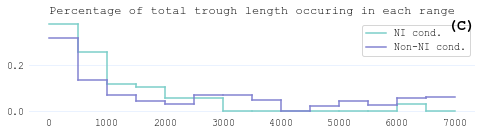}
    \includegraphics[width=13cm]{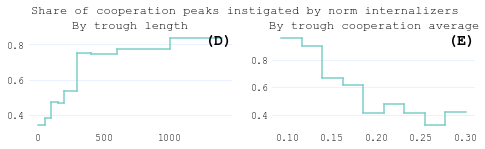}
    \caption{\textbf{Troughs characteristics in the naturalistic model.} \textbf{(A), (B)} The intervals of above-average cooperation (the spikes) are shown as dark lines for each of the simulation runs. While trough lengths for the NI-condition (right) are, on average, slightly longer, it is clear that the non-NI condition (left) has a fatter tail of very long troughs, which can be observed more quantitatively in \textbf{(C)}. This histogram plots the probability of being in a trough of various lengths, given that one is in a trough (longer troughs therefore have higher weight since a randomly sampled round is more likely to be in a long than a short trough). \textbf{(D), (E)} Plot the share of cooperation spikes \emph{initiated} by norm internalizers after troughs of various lengths and depths, showing that norm internalizers tend to initiate the spikes that follow especially \emph{long} and \emph{low} troughs. Figures use \texttt{default} parameters. Corresponding figures for the abstract model are in \ref{appendix:counterpart-figures}; they do not exhibit the same pattern}
    \label{fig:spike-trough-comparison}
\end{figure}
\end{center}

\begin{center}
    \begin{figure}[htbp]
    \centering
    \centerline{\includegraphics[width=15cm]{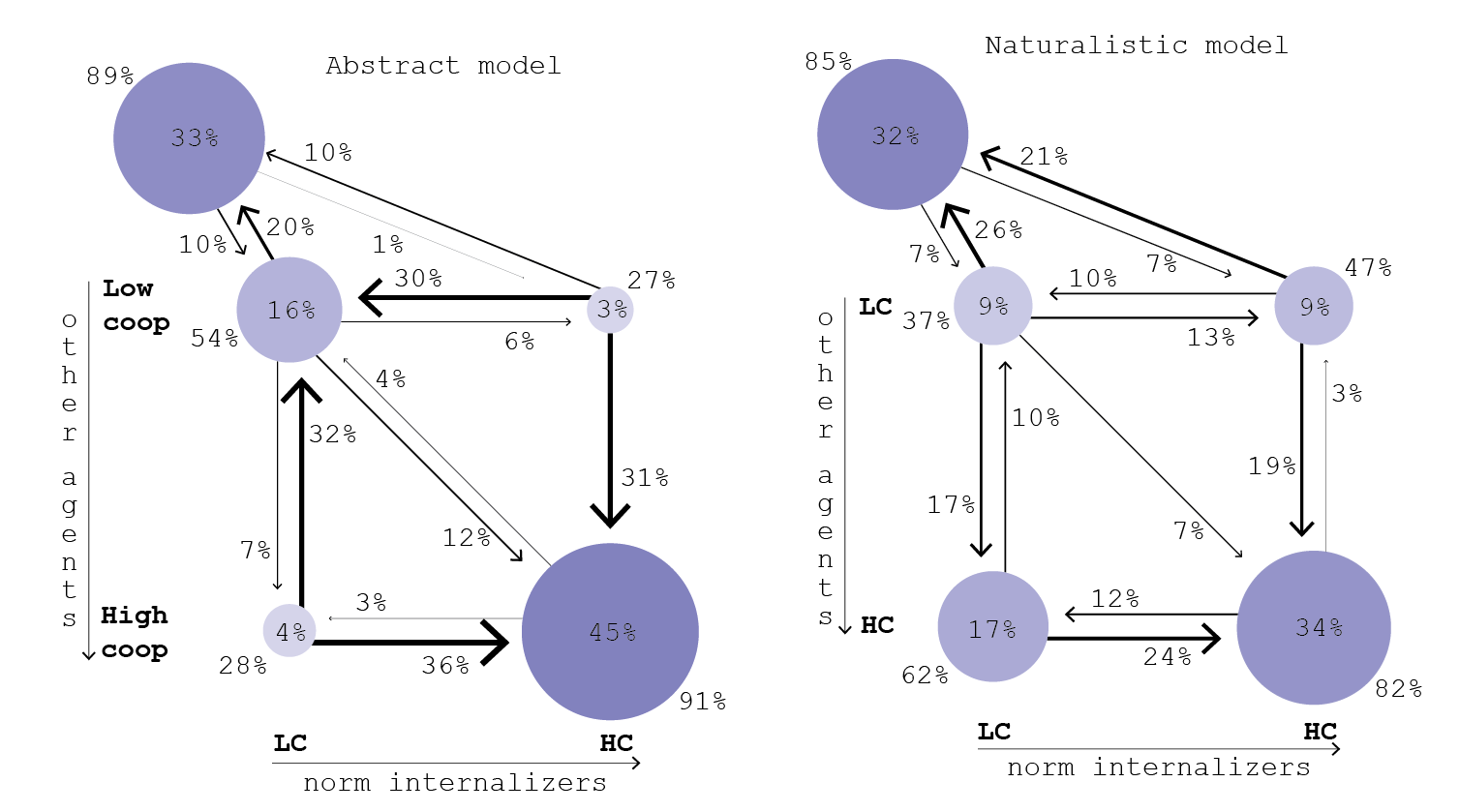}}
    \caption{\textbf{Transitions in the abstract (left) and naturalistic (right) models.} Both diagrams represent state transitions in the norm internalizer (NI) condition. The vertical axis shows L(ow)/H(igh) cooperation of non-NIs, the horizontal axis L/H cooperation for NIs. The top left circles represent low cooperation for both, in the special case when the cooperation is more than half a standard deviation below the mean. The arrows shown are the top two outgoing arrows for each node, or all arrows above 5\%, whichever set is larger. Circle sizes, along with the percentage listed inside, represent the share of rounds spent in each state; circle tint, along with the percentage listed outside, represents the percentage of self-loops. Each round in the naturalistic (abstract) model is divided into disjoint segments of 25 (10) rounds; transition probabilities are calculated from one window to the next. \texttt{Default} parameters were used}
    \label{fig:state-transitions}
\end{figure}
\end{center}

As can be seen in Fig.~\ref{fig:spike-trough-comparison}, there are more long troughs in the non-NI condition. For the NI condition, if a point is in a trough, there is a 3\% probability that it is a trough of more than 3,000 rounds, while for the non-NI condition, that number is 33\%. We sought to elucidate this result by looking at which agent type \emph{initiated} cooperation spikes -- that is, the agent type that exceeded its mean cooperation level first. This result is plotted in Fig.~\ref{fig:spike-trough-comparison}(d, e), which shows that for both longer and deeper troughs, NIs played a disproportionate role in initiating subsequent spikes. This helped prevent the NI-condition runs from getting stuck in very long troughs.
A useful way to examine this phenomenon is with a state-transition diagram, shown in Fig.~\ref{fig:state-transitions}. While the transitions between these particular states are not technically Markovian, they nonetheless show the likely successor for each state. There are two things to observe here. First, the most stable states (the ones with the greatest proportion of self loops) are those with very low cooperation (top left), and those with above average cooperation of both NIs \emph{and} non-NIs (bottom right). Effective `catalysis' of cooperation, then, corresponds to moving from the top left to the bottom right. Notice that, in the naturalistic model, one of the most likely paths that achieves this is the one from very low cooperation (upper left) to high-NI cooperation (top right) to high cooperation for all agents (bottom right). All other two- or three- step paths have a much lower probability. The advantage of this path is much less pronounced in the abstract model (left), which is consistent with our observation that NIs don't play as much of a catalyst role in that setting.

\subsubsection{Stabilize}

While the catalyst role was much stronger in the naturalistic model, the third, stabilizing type of effect that NIs had on high levels of cooperation (reflected in the longer peak lengths in the NI condition, along with the higher average levels of cooperation) was strongly present in both. This may seem counter-intuitive, because the peaks of NIs themselves are shorter than those of other agents. The key point, however, is that when the NI population is higher than average, cooperation peaks among \emph{all} agent types, not just NIs, tends to be maintained (right side of Fig.~\ref{fig:spike-summary}.

\begin{center}
    \begin{figure}[htbp]
        \centering
        \centerline{\includegraphics[width=16cm]{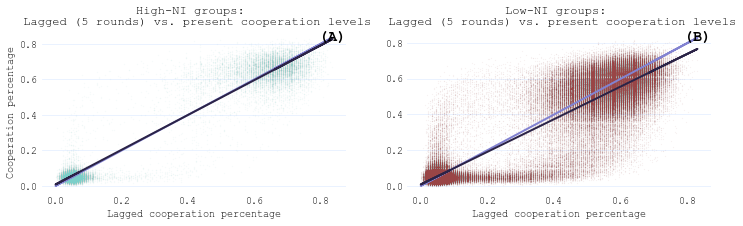}}
    \caption{\textbf{Norm internalizers as stabilizers.} Both figures show the level of cooperation in a given round vs.\ five rounds earlier, for high-NI ($>$20\%) \textbf{(A)} and low-NI  ($<$20\%)\textbf{(B)} groups in the abstract model. The best-fit (dark) is much closer to the slope-1 line (light) in high-NI groups, with coefficients of 0.91 ($\pm$ 0.0015) vs 0.97 ($\pm$ 0.003), respectively (and approximately equal constants). This may not seem like a huge difference, but iterating a cooperation level of 1 through this linear function requires 82 cycles to fall below $\frac{1}{2}$ in high-NI case, while in the low-NI case, it only requires 8 cycles. \texttt{Default} parameters were used; confidence intervals are 95\% two-tailed $t$-test. Equivalent figures for the naturalistic model, showing the same pattern, are in Appendix~\ref{appendix:counterpart-figures}}
    \label{fig:lin-reg-stabilizers}
\end{figure}
\end{center}

\begin{center}
    \begin{figure}[htbp]
    \centering
    \begin{subfigure}{.5\textwidth}
        \centering
        \raisebox{6mm}{
        \includegraphics[width=7.5cm]{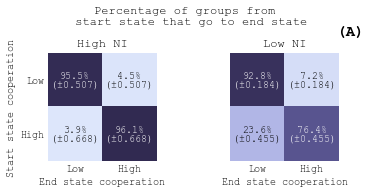}}
    \end{subfigure}%
    \begin{subfigure}{.5\textwidth}
      \centering
      \includegraphics[width=7.5cm]{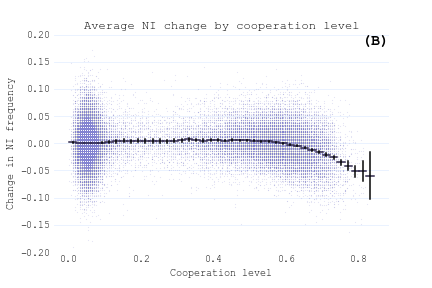}
    \end{subfigure}
    \caption{\textbf{Norm internalizers (NIs) as stabilizers, and their shortcoming.} \textbf{(A)} The probability of transitioning from a state of high ($>$50\%) to low ($<$50\%) cooperation for groups that have high ($>$20\%) or low ($<$20\%) proportions of NIs, in the abstract model. High-NI groups maintain high cooperation dramatically better than low-NI groups (though they have a slightly harder time going from low to high cooperation). \textbf{(B)} The flaw: here we plot the difference in NI-population at different cooperation levels in the abstract model. Norm internalizer populations tend to fall when cooperation gets high. Both figures show error bars and confidence intervals of 99.9\% ($z$-test), and \texttt{default} parameters. For equivalents in the naturalistic model, exhibiting the same pattern, see Appendix~\ref{appendix:counterpart-figures}}
    \label{fig:stabilizer-and-flaw}
\end{figure}
\end{center}

We examine this effect by zeroing in on individual groups and their cooperation patterns. In particular, we first examine the regression coefficient between past and future levels of cooperation in high-NI vs.\ low-NI groups. We found that the coefficient is much closer to $1$ in high-NI groups, which implies a much better sticking power of cooperation (see Fig.~\ref{fig:lin-reg-stabilizers}). We also examined state transitions between low ($<$50\%) and high ($>$50\%) cooperation, finding that a high level of NIs was much more effective at maintaining high cooperation levels: 96\% of the time, high cooperation groups stayed high in high-NI groups, but only 76\% of the time in low-NI groups (see Fig.~\ref{fig:stabilizer-and-flaw}a). 

While high populations of NIs are effective at maintaining high levels of cooperation, it is also true that high levels of cooperation lead to a decline in their population (see Fig.~\ref{fig:stabilizer-and-flaw}b). A Granger causality test of the effect of cooperation levels on norm internalizer population supported this observation: with very high confidence ($p < 0.0001$), high cooperation levels Granger-caused lower subsequent NI populations. This echoes the tendency for NI populations to fall at higher b:c ratios (Fig.~\ref{fig:norm-internalizer-quick-stats}b), when cooperation levels tend to be higher. Furthermore, while NIs tended to help high cooperation groups stay high, they also had a small but significant role in keeping low-  (under 50\%) cooperation groups low (see Fig.~\ref{fig:stabilizer-and-flaw}a). We will discuss the apparent tension between this fact and the catalyst role of norm internalization in 
Section~\ref{section:discussion-catalyze}.

\section{Discussion} \label{section:discussion}

The previous section highlighted the evidence for the roles that norm internalizers (NIs) played in raising average levels of cooperation. Here, we offer mechanistic explanations (or hypotheses), and ease some of the apparent tensions between the three roles. 

\subsection{Overall effect}

First, we examine the overall effect of NIs on cooperation levels: when they were present \emph{in the strategy space} --- not necessarily when their population was larger --- they tended to boost cooperation levels (Fig.~\ref{fig:summary}). This is consistent with the result that, when selection on a trait depended on Darwinian fitness (as the helping trait does in our model), unbiased imitation can help increase cooperation levels \cite{lehmannCulturalTransmissionCan2008}. The difference is, in our model, norm internalization is not an exogenous given but an evolving trait that can go extinct. In this latter respect our work is similar to \cite{lehmannCoevolutionCulturallyInherited2008}, though we obtained different results likely because their model included neither persistent groups (they reshuffled every round) nor punishment. For NIs to have any effect in our model, both of these mechanisms were required. This echoes Gavrilets and Richerson's finding about \emph{non-conformist} NIs \cite{gavriletsCollectiveActionEvolution2017} they, too, had a much higher effect on behavior in the presence of punishment.

Furthermore, most of that boost came not from NIs own higher cooperation, but from increased cooperation levels of other agents: even if NIs \emph{did} cooperate more than other agents (which was not always the case), it only made a small difference in the overall mean since their population was so small.  Importantly, the three roles of NIs discussed earlier do not require higher long-term average levels of cooperation among NIs: instead, they effectively carry other agents along. As catalysts and stabilizers, NIs raised or maintained cooperation levels of other agents. And as polarizers, they adopted (a more extreme version of) whatever agents in their group were already doing.

How do NIs carry other agents along? In our model, there is a basic mechanism accomplishing this. Those caught defecting are punished by an indirect reciprocity mechanism: the group benefit of cooperation is withheld from them. Notice that the group benefit increases as the number of cooperators in the group increases, so it becomes more and more costly to defect as cooperation becomes more common. This increasing cost of defection causes \emph{other} agents to cooperate more often as cooperation becomes more common, even if they are not NIs. However, while the payoff difference between cooperation and defection decreases when cooperation becomes common, the individual payoff of cooperating rarely, if ever, exceeds the cost of defection. This contrasts with the structure of a stag hunt game, in which it pays to simply adopt the majority strategy.

\subsubsection{Population-cooperation tradeoff}\label{section:discussion-population-cooperation}

An important factor in the ability of the NI-condition to catalyze cooperation is the trade-off between population and cooperation level. The NI population has to be high enough to make a difference, but that is not enough: it is of no help for them to constitute a third of the population if their cooperation level is kept down as a result of internalizing the norm of defection (this is what happened, for instance, at a benefit-to-cost ratio (b:c) of \texttt{3.0} in the abstract model). But --- and here is the catch --- the more NIs cooperate, the more their population falls (Fig.~\ref{fig:stabilizer-and-flaw}). Thus, to raise cooperation levels, NIs must cooperate, but so much that their population falls. The presence of this tradeoff explains why NIs are most effective with mid-range b:c ratios: if b:c is too high, NIs cooperate too much, leading their population to diminish. If it is too low, NIs never internalize the norm of cooperating, so their numbers can grow to make up about a third of the population without helping to spark cooperation (see Fig.~\ref{fig:norm-internalizer-quick-stats}). These fluctuating population dynamics are in stark contrast to the results of Gavrilets and Richerson, who observe a more or less stable level of norm internalizers once the trait has caught on \cite{gavriletsCollectiveActionEvolution2017} This is no surprise, as their version of norm internalization is not a conformist one, and should therefore be less sensitive to the population distribution of other cooperators. 

The mid-range is  especially important as the region where cooperation is possible, but the hardest to maintain. One may of course ask, is the added cognitive machinery required for the norm internalization strategy worth the cost if all it does is lower the threshold for appreciable amounts of cooperation to a lower b:c ratio? The answer is quite possibly yes, especially because the machinery required to imitate the behaviors of others may have already existed for copying beneficial behaviors, and could have been co-opted at minimal cost (see e.g.\ \cite{henrichEvolutionConformistTransmission1998, gintisHitchhikerGuideAltruism2003}).

While the naturalistic and abstract model yielded largely similar results, there were some differences, including two major qualitative differences. First, in the abstract model, norm internalizers always cooperated more than others, while in the naturalistic model, it depended very much on the b:c ratio. Second, the NI-condition in the abstract model actually saw increasing or flat cooperation levels as migration rates increased, compared with declining ones in the spatial model. While we cannot say definitively what caused these differences, it makes sense that migration had different effects in the two models, as the spatial structure of the naturalistic model meant that only agents from neighboring groups mixed, while in the abstract model, the mixing was uniform.

We next discuss the three roles played by norm internalizers.

\subsection{Polarize}

We mentioned in Section~\ref{section:introduction-group-selection} that between-group differences are necessary for group selection to be strong. Norm internalization, by polarizing groups, increases between-group differences, which in turn can enhance selection in favor of cooperation. Our result is thus in line with previous results on conformist transmission enhancing inter-group differences \cite{henrichCulturalGroupSelection2004}.

There may seem to be a tension between two findings: high-NI groups do not have higher average cooperation (Fig.~\ref{fig:polarize}), but nevertheless, the NI condition has higher cooperation than the non-NI condition. However, even though NIs may have an ambiguous effect on intra-group cooperation, the differences they cause \emph{between groups} can enhance group selection in favor of cooperation. We will discuss in the next section (\ref{section:discussion-catalyze}) how the inter-group dynamics resulting from polarization could lead to catalysis, and then in the following one, (\ref{section:discussion-stabilize}), how the intra-group effects of polarization can lead to stabilization. Hence, polarization is the central role whose different manifestations help give rise to the others.

\subsection{Catalyze} \label{section:discussion-catalyze}

In their role as catalyst, NIs helped to spark spikes in cooperation when cooperation levels were especially low. In this section, we propose a mechanism by which this might have occurred, resolve apparent tensions between this function and the other two, and provide hypotheses for why this role was less salient in the abstract model. 

As can be seen in Fig.~\ref{fig:stabilizer-and-flaw}, NIs face no selective disadvantage at low levels of cooperation. This means that when cooperation is very low, their population can grow quite large, by drift. Of course, we know that a large NI population is not sufficient to bring about cooperation, as they can internalize a defection norm. So why might this population start cooperating? The answer is that even in global states of low cooperation, there will be variation between groups: some will cooperate more, others less. And crucially --- due to the polarizing tendency (here is the promised link between polarization and catalysis), and the higher variability of NI cooperation  (Fig.~\ref{fig:norm-internalizer-quick-stats}) --- the groups with unusually high or unusually low cooperation will probably be the ones with more NIs. Especially in environments where cooperation is globally very low, the higher variance caused by NIs is needed to produce higher-cooperation groups. These groups are favored by group selection, and their spread is what brings about a global spike in cooperation. Notice that catalysis is thus a result of the \emph{inter-group} dynamics resulting from polarization.

Why, then, was catalysis less pronounced in the abstract model? The catalysis mechanism that we propose requires \emph{simultaneous} inter-group variation: in the same round, there must be some higher and some lower cooperation groups. This may be much easier to achieve when there is spatial structure: the high migration rates means that groups mix quite quickly, but the spatial structure in the naturalistic model allows for geographically structured inter-group variation that is not possible in the abstract model. 

This proposed mechanism helps to resolve apparent tensions between the catalyst role and the other two. One might wonder, in particular, how NIs can be catalysts of cooperation when Fig.~\ref{fig:stabilizer-and-flaw}a, demonstrating stabilization, shows that it is harder for high-NI groups to go from low cooperation to high cooperation (due to the polarizing tendency of norm internalization bringing groups under 50\% cooperation closer to zero). The key insight is that the figure shows the \emph{intra-group} effect of NIs: our proposed mechanism recognizes that intra-group, NIs have an ambiguous effect on cooperation, and it is only when we consider the \emph{inter-group} variation, along with group selection favoring cooperative groups, that the catalyst effect can work. 

\subsection{Stabilize} \label{section:discussion-stabilize}

The stabilizing mechanism is more intuitive, and has two parts. First, when NIs are present in larger quantities, and cooperation is high, they will be internalizing the cooperation norm --- this is simply the positive side of the polarization phenomenon. This will further increase the levels of cooperation, causing the peaks of high cooperation to stick longer. Second, as described earlier, the increases in NI cooperation can raise defection costs, causing other agents to cooperate more. Note that the mechanism proposed here is the result of the \emph{intra-group} consequences of polarization. This stabilization effect is in line with Henrich and Muthukrishna's idea that group selection is greatly aided by intra-group dynamics that help to maintain a particular state (in this case, high cooperation) \cite{henrichOriginsPsychologyHuman2021}.

Why, given the positive-feedback dynamics described so far, doesn't cooperation persist indefinitely? The answer is that the positive feedback loop is embedded in a larger negative feedback loop. It is true that high cooperation leads to more internalization of cooperation and a higher cost of defection, which in turn lead to more cooperation; but high cooperation levels also tend to reduce the population of NIs, as seen in Fig.~\ref{fig:stabilizer-and-flaw} (the same effect leads to a reduction in NI population with rising b:c, as described in Section~\ref{section:discussion-population-cooperation}). This is because when cooperation levels are very high, NIs become unconditional cooperators, and cooperate much more than other agent types, leading to a reduction in their population. This results in a subsequent reduction in cooperation levels, but if cooperation levels get too low, NI populations can rise once more, setting off the positive feedback of cooperation again.

The cyclical dynamics can also help explain how NIs prolong overall peak durations while having shorter peaks themselves. When NI cooperation levels rise, they bring the cooperation of other agents up, too. However, in groups where cooperation gets too high, norm internalizer populations tend to die off; globally, this has the effect of killing the highly-cooperative NIs and preserving the non-cooperative NIs, which reduces the NI cooperation average, ending the NI cooperation peak. However, cooperation levels of other agent types, which had grown in response to an increasing cost of defection, remain high because their own high cooperation levels maintain that high cost of defection --- so the cooperation peak of other agents persists. When it eventually does fall enough (though remaining relatively high), it opens the door for norm internalizer populations to rise again, and possibly start to bring cooperation levels back up. In this way, NI cooperation can spike when it is needed to preserve a peak of non-NI cooperation, but it doesn't necessarily remain high for the entire non-NI peak, thereby increasing peak lengths even though its own peak lengths are comparatively short.

\subsection{Cultural Evolution}

So far we discussed the immediate effects of norm internalization on cooperation dynamics. But there is also a broader consequence to its effects, namely, that it forms part of the basis for cultural evolution. Under genetic evolution, transmission occurs from parent to offspring, with variation introduced by mutation and recombination. Cultural evolution, in comparison, requires that individuals in a society imitate others, for example, through conformist transmission.

Existing explanations for conformist transmission \cite{gintisHitchhikerGuideAltruism2003} do not explain why individuals would copy obviously self-sacrificial behaviors (see Section~\ref{section:background-possible-causes}), short of ``myopia'' or inability to tell the difference. The present research provides an alternative explanation for conformist transmission in these hard-to-explain cases. Here, group selection plays an important role, amplified by the ability of norm internalization to increase differences between groups. Our research, therefore, helps elucidate why humans didn't evolve into the fabled \textit{Homo economicus} \cite{henrichSearchHomoEconomicus2001}, and provides grounding for a broader theory of cultural transmission of altruistic and moral behaviors.

\section{Conclusion} \label{section:conclusion}

In this paper, we looked at how the presence of norm internalization in the strategy space was able to increase cooperation levels, beyond what group selection and punishment by indirect reciprocity were able to do alone. This was so even though norm internalizers (NIs) tended to make up a small portion of the population, and they didn't necessarily cooperate more than other agent types. 

We were motivated by the question of whether NIs might be able to fill in the gap by a certain proportion of actions (about half) being unobserved, and therefore, unpunished. The answer was yes, but their effect was mostly indirect: a minority of NIs increased the level of cooperation of all strategies. This accords with our intuitions about the role of \emph{conscience} --- conscience is essentially norm internalization paired with emotions like guilt and shame to enforce norm-following `from the inside'. Our intuition suggests that a conscience-like mechanism could help fill the gap when punishment is imperfect, and indeed, our results bear that out. 

We identified three related roles that NIs played: they \emph{polarized} groups, strengthening group selective forces in favor of cooperation, they \emph{catalyzed} cooperation when global levels were especially low, and they \emph{stabilized} bouts of high cooperation, keeping them going for longer.

We showed these roles in action, and, in Section~\ref{section:discussion}, we proposed some more mechanistic explanations for each. Polarization was the hub role, which led to catalysis through \emph{inter-group} dynamics and stabilization through \emph{intra-group} dynamics. 

For our inter-group dynamics, we used a wide range of migration parameters, and where required, empirically-backed conflict parameters. We did not address the question of how punishment by indirect reciprocity might remain stable; for this, see \cite{ohtsukiLeadingEightSocial2006, odouardTitTattlingCooperation2023}.

Importantly, we tested the effect of NIs over and above the effects of group selection and punishment, by looking at two conditions: one in which NIs were present in the strategy space, and one in which they were not. Given that both these forces are at play, a strategy space that includes norm internalization yields higher cooperation than a strategy space that only has unconditional cooperators, and it can do so when both cooperation and norm internalization start out rare. As we have shown, they remain relatively uncommon, yet facilitate a dynamic that leads other agents to cooperate more and for longer.

Our work may be extended in several promising directions:

\begin{itemize}
\item Morality consists of much more than just cooperation in public goods games. Coordination games, hawk-dove games, and stag hunts all play into morality \cite{curryMoralityCooperationProblemcentred2016, curryItGoodCooperate2019}. Further exploration might aim to understand the effects that norm internalization would have in each of these diverse contexts.
\item While we used a continuous strategy space, it was still restricted to strategies that employed a particular function of certain input variables. Allowing for strategy optimization, for instance by making the agents capable of reinforcement learning, is a particularly interesting direction for future work.
\end{itemize}

\begin{center}
    \textsc{Acknowledgments}
\end{center}
We would like to thank Michael Macy, Alex Vladimirsky, Michael Holton Price, Tyler Millhouse, and David Krakauer for their thoughts, ideas, and feedback; our anonymous reviewers, whose thorough comments allowed us to greatly improve the paper; and David Krakauer and the Santa Fe Institute, under whose auspices we were able to complete the last painstaking steps of this research.

\begin{center}
    \textsc{Funding}
\end{center}
This work is supported by a grant to SFI from the Omidyar Network to cover core research in the area of Emergent Political Economy. David Krakauer is
the PI on this award. The funders played no role in the design, execution, writing, and submission of this research. 

\begin{center}
    \textsc{Conflicts of interest}
\end{center}
We have no conflicts of interest to disclose.

\begin{center}
        \textsc{Data Availability Statement}
\end{center}
The datasets generated and analyzed during the current study are available on Google Drive at \texttt{\url{https://drive.google.com/drive/folders/191NgPRAGVb0q4hbv9BUPXqfh7lSLAKpv?usp=sharing}}. Furthermore, the full codebase for the models, including the code we used to analyze the data, can be found at \texttt{\url{https://github.com/victorvikram/norm-internalization-and-coop}}.


\newpage
\bibliographystyle{apalike}
\bibliography{bibliography}

\newpage  
\appendix
\singlespacing
\gdef\thesection{Appendix \Alph{section}}
\section{Model: Full(er) specification} \label{appendix:model-fuller-specification}
\subsection{Abstract model} \label{appendix:abstract-model}

The parameters for the abstract model are shown in Table~\ref{table:full-abstract-parameters}.

\begin{center}
   \begin{tabular}{c c p{10cm}}
        Category & Variable & Description  \\
        \hline
        \multirow{5}{*}{\textit{Game}} & $b$ & the benefit produced from cooperating \\
        & $c$ & the cost of cooperating \\
        & $\bar{b}$ & the proportion of the public pot that each (unpunished) agent received in the previous round \\
        & $f$ & the base fitness, which all agents receive in every round \\
        & $p_{\text{obs}}$ & the probability of being observed \\
        \hline
        \multirow{5}{*}{\textit{Strategies}} & $\epsilon$ & the probability of acting counter to one's strategy \\
        & $\beta$ & how much the present value is weighted into a running average \\
        & $l$ & the learning rate \\
        & $d$ & the starting distribution of strategies \\
        & $d_{\text{mut}}$ & the distribution on strategies when a mutation occurs \\
        \hline 
        \multirow{6}{*}{\textit{Environment}} & $p_{\text{con}}$ & The probability that a given group will engage in a conflict \\
        & $p_{\text{mig}}$ & probability of migration \\
        & $p_{\text{mut}}$ & the probability of offspring mutating to another strategy \\
        & $\sigma$ & the survival rate \\
        & $n$ & number of agents in a group \\
        & $g$ & number of groups \\
        & $y$ & number of rounds to run the model for
    \end{tabular} 
    \captionof{table}{\textbf{Parameters of the abstract model}}
    \label{table:full-abstract-parameters}
\end{center}

\subsubsection{Decision} \label{appendix:abstract-decision}
When deciding whether to cooperate, all agents use the same basic inequality, given in Equation~\ref{eqn:cooperation-inequality-spatial}:

\begin{equation}
p\overline{b} \geq (1 - \pi)c,
\end{equation}
though if it is the first round, they cooperate if their propensity to cooperate is greater than $1/2$, and defect otherwise. Like in the abstract model, agents will deviate from their strategy with probability $\epsilon$.

Agents learn their parameter $\pi$ in different ways.

\begin{itemize}[leftmargin=0.75in]
    \item[\textit{Selfish}]  After each round, they compare their payoff $w$ to their present-biased average payoff $\bar{w}$, and their current propensity to cooperate $\pi$ to their present-weighted average $\overline{\pi}$ (the present-weighted averages are calculated using $\beta$ to determine the weight of the present value). They then increment $\pi$ according to
    \begin{equation}
        \pi \leftarrow \pi + \frac{(\pi - \overline{\pi})(w - \overline{w})}{\overline{w} + w}.
    \end{equation}
    Thus, if their fitness was above average and their $\pi$ was above average, $\pi$ will increase, if their fitness was below average and $\pi$ was above average, $\pi$ will decrease, and so on. The denominator is present to normalize the increment.
    \item[\textit{Norm internalizer}] The norm internalizer is learning strategy is conditional on the number of other cooperators. If the more than a majority are cooperating, the update is
    \begin{equation}
        \pi \leftarrow (1 - l)\pi + l,
    \end{equation}
    otherwise, it is
    \begin{equation}
        \pi \leftarrow (1 - l)\pi.
    \end{equation}
    These should be interpreted as the weighted average of $\pi$ and 1 (moving closer to an unconditional cooperator) and the weighted average of $\pi$ and 0 (moving closer to an unconditional defector). The weight is given by the learning rate, $l$.
    \item[\textit{Static}] - this strategy keeps its $\pi$-value constant.
\end{itemize}

\subsubsection{Intergroup dynamics} \label{appendix:abstract-intergroup}
In every round, groups are paired up at random, and with probability $p_{\text{con}}$ they will engage in a conflict. If groups $i$ and $j$ fight, the probability that $i$ will win is given by 
\begin{equation}
    \dfrac{1}{2}(1 + \dfrac{\overline{w_i} - \overline{w_j}}{\overline{w_i} + \overline{w_j}}), 
\end{equation}
where $\overline{w_i}$ is the average fitness of group $i$. The victor then replaces the old group with a copy of itself. Thus, higher fitness groups have an advantage in group conflict.

\subsubsection{Selection} \label{appendix:abstract-selection}
Agents act according to their strategy and receive their payoffs according to the rules of the game. Then, in every round, there is a phase of mortality. A survival rate $\sigma$ determines the proportion of agents in each group that will survive at the end of each round. A set of $n \cdot \sigma$ agents are selected (without replacement) to survive, where an agent is selected to survive with probability
\begin{equation}
    \dfrac{w}{n\overline{w}},
\end{equation}
where $w$ is the agent's fitness and $\overline{w}$ is the average fitness of the group. 

Each time an agent dies, another agent in the group is chosen to reproduce, where the probability of reproduction is proportional to that agent's fitness. Thus, for every death, any given agent with fitness $w$ is selected to reproduce with probability $\dfrac{w}{n\overline{w}}$. With probability $(1 - p_{\text{mut}})$, the new agent has the same learning style as their parent. Otherwise, the new agent's strategy is chosen at uniformly at random among the learning styles in the strategy space. When norm internalizers are not present in the strategy space, their probability weight is allocated towards unconditional cooperators.

\subsection{Naturalistic model} \label{appendix:naturalistic-model}
The parameters for the naturalistic model are shown in Table~\ref{table:naturalistic-parameters}.
\begin{center}
   \begin{tabular}{c c p{10cm}}
        Category & Variable & Description  \\
        \hline
        \multirow{7}{*}{\textit{Game}} & $b$ & the benefit from cooperation* \\
        & $c$ & the cost of cooperating* \\
        & $c_{\text{dist}}$ & the cost of foraging on an adjacent square (rather than one's own)* \\
        & $f$ & the base fitness, a constant proportional to the resources on a square* \\
        & $\bar{b}$ & the proportion of the public pot that each (unpunished) agent received in the previous round\\
        & $p_{\text{obs}}$ & the probability of being observed \\
        & $c_{\text{alive}}$ & the cost of staying alive for another round \\ 
        & $c_{\text{repro}}$ & the cost of reproducing \\ 
        \hline
        \multirow{6}{*}{\textit{Strategies}} & $\epsilon$ & the probability of acting counter to one's strategy \\
        & $\beta$ & how much the present value is weighted into the average \\
        & $l$ & the learning rate \\
        & $d$ & the starting distribution of strategies \\
        & $d_{\text{mut}}$ & the distribution on strategies when a mutation occurs \\
        \hline 
        \multirow{5}{*}{\textit{Environment}}
        & $p_{\text{mut}}$ & the probability of offspring mutating to another strategy \\
        & $s$ & grid side length \\
        & $n$ & starting number of agents in a group \\
        & $g$ & starting number of groups \\
        & $y$ & number of rounds to run the model for
    \end{tabular}
    \captionof{table}{\textbf{Parameters of the naturalistic model} *strictly speaking, these quantities are not the final costs and benefits; they are first divided by the number of agents foraging on the square} \label{table:naturalistic-parameters}
\end{center}

\subsubsection{Decision}\label{appendix:naturalistic-decision}
For the decision on whether to cooperate, the naturalistic model functions exactly as the abstract model (\ref{appendix:abstract-decision}). As for the decision on where to forage, all agents use the same strategy. First, they calculate expected payoffs of staying on their current square versus going to an adjacent square as follows:
\begin{equation}
    \begin{aligned}
    \pi_{\text{stay}} &= \frac{f}{n_{\text{here}}} \\
    \pi_{\text{go}} &= \frac{f - c_{\text{dist}}}{n_{\text{out}}}
    \end{aligned}
\end{equation}
where $n_{\text{here}}$ is the number of agents on the current square and $n_{\text{out}}$ is the number on adjacent squares. Also, $\pi_{\text{go}}$ factors in the cost of foraging on distant lands. Using these payoffs, the agent adopts probabilities for staying or going as follows:
\begin{equation}
    \begin{aligned}
    p_{\text{stay}} &= \frac{\pi_{\text{stay}}}{\pi_{\text{stay}} + \pi_{\text{go}}}  \\
    p_{\text{go}} &= \frac{\pi_{\text{go}}}{\pi_{\text{stay}} + \pi_{\text{go}}}
    \end{aligned}
\end{equation}
Agents decide whether to stay or to go based on these probabilities. If they decide to go, they choose one of the four adjacent squares at random.

\subsubsection{Distribution}\label{appendix:naturalistic-distribution}
The base-fitness payoff (interpreted as the payoff from simply foraging alone) is $f/n_{\text{here}}$, to account for the dwindling of resources as the square gets more crowded. Cooperating will produce a benefit to the group of $b/n_{\text{here}}$ because cooperation bears more fruit when fewer agents compete for resources. The cost paid to cooperate is $c/n_{\text{here}}$ and the cost paid to forage on distant lands is $c_{\text{dist}}/n_{\text{here}}$ (where $n_{\text{here}}$ is the number of agents on the new square). Both of these costs are inversely proportional to the number of agents because, fundamentally, both of these are costs in \textit{time}. If the resources on a particular square are rich (that is, there are few competitors for the resources), then a single unit of time is worth more, because one unit of time foraging will bear more fruit. Thus, if agents pay a constant cost in \textit{time} either to cooperate or to forage on distant lands, the fitness cost will be proportional to the richness of the resources on their square (and inversely proportional to the number of agents on their square). 

Thus, an agent generates value shown in Table~\ref{table:naturalistic-generation}, but because they do not keep the group benefit component that they generate (these get totaled up at distributed to all the agents in the group who were not caught defecting) the final payoff is obtained by subtracting the $b$ wherever it appears and adding back $\overline{b}$ to any agent that was not caught defecting.

\begin{center}
   \begin{tabular}{c c c}
        & Defect & Cooperate  \\
        \hline
        Here & $\dfrac{f}{n}$ & $\dfrac{f-c}{n} + b$ \\[0.6em]
        \hline
        Distant & $\dfrac{f - c_\text{dist}}{n}$ & $\dfrac{f-c - c_\text{dist}}{n} + b$ \\
    \end{tabular}
    \captionof{table}{\textbf{Value generated by agents according to their decisions} This is not their final payoff, because it is pre-distribution.}\label{table:naturalistic-generation}
\end{center}

\subsubsection{Intergroup dynamics}\label{appendix:migration-remark}
Here we make a remark about how the migration rates in the abstract and naturalistic models. They are not commensurate, for two reasons. The first is that lifetimes in the naturalistic model are much longer (20 as opposed to 10/3 rounds), so a certain lifetime migration rate in the naturalistic model would require a lower round-by-round migration rate. But, to complicate matters, agents in the naturalistic model only migrate if they are foraging on another square where another group is located. The migration probability in this very specific situation is the parameter $p_\text{mig}$. This would suggest that the naturalistic model would require a \emph{higher} migration rate than the abstract model for the same effect. In the end, for the abstract model, we determined that a lifetime migration rate of 50\% would require a $p_\text{mig} = 0.2$. For the naturalistic model, we empirically determined that $p_\text{mig} = 0.5$ leads to approximately a 50\% migration rate.

\subsubsection{Selection} \label{appendix:naturalistic-selection}
Every round, an agents' fitness score gets incremented by their payoff. Each agent must then  pay $c_{\text{alive}}$ (the cost of stayin' alive), and if they cannot, they die. Further, every agent has a randomly sampled lifespan, and when they exceed their lifespan, they die. 

If an agent can afford to pay the $c_{\text{repro}}$, the reproduction cost, and still have some fitness left over (equal to $c_{\text{alive}}$), they will reproduce. The child is a noisy copy of their parent, with the same learning style and propensity to cooperate, but some noise added to the propensity to cooperate, and probability $p_{\text{mut}}$ of a mutation in learning style (in which case one of the three are chosen at random).

\section{Results: Parameters} \label{appendix:results-parameters}
In all tables in this section, we will refer to two values of $d$ and $d_\text{mut}$: \texttt{non-NI} (no norm internalizers), \texttt{NI} (with norm internalizers). These names correspond respectively to distribution ($d$) values of

\noindent
\texttt{selfish: 0.49 \newline static: 0.49 \newline norm internalizer: 0.02,}
\newline \newline \noindent
\texttt{selfish: 0.49 \newline static: 0.49 \newline uncond-coop: 0.02},
\newline \newline \noindent
and mutation distributions ($d_\text{mut}$) of 
\newline \newline \noindent
\texttt{selfish: 1/3 \newline static: 1/3 \newline norm internalizer: 1/3,}
\newline \newline \noindent
\texttt{selfish: 1/3 \newline static: 1/3 \newline uncond-coop: 1/3}.
\newline \newline \noindent

\subsection{Abstract Model }\label{appendix:abstract-model-parameters}

\begin{center}
   \begin{tabular}{c c p{10cm}}
        Category & Variable & Description  \\
        \hline
        \multirow{3}{*}{\textit{Game}} & $b$ & \texttt{3.0, 3.25, 3.5, 3.75, 4.0, 4.25, 4.5} \\
        & $c$ & \texttt{1} \\
        & $f$ & \texttt{1} \\
        \hline
        \multirow{3}{*}{\textit{Strategies}} 
        & $\epsilon$ & \texttt{1/20} \\
        & $\beta$ & \texttt{0.2} \\
        & $l$ & \texttt{1/2} \\
        & $d$, $d_{\text{mut}}$ & \texttt{non-NI}, \texttt{NI} \\
        \hline 
        \multirow{7}{*}{\textit{Environment}} & $p_{\text{con}}$ & \texttt{1/13} \\
        & $p_{\text{mig}}$ & \texttt{0, 0.1, 0.2, 0.3, 0.4, 0.5, 0.6} \\
        & $p_{\text{mut}}$ & \texttt{1/100} \\
        & $\sigma$ & \texttt{7/10} \\
        & $n$ & \texttt{35} \\
        & $g$ & \texttt{60} \\ 
        & $y$ & \texttt{15,000}
    \end{tabular} 
    \captionof{table}{\textbf{Parameters of the abstract model}}
    \label{table:abstract-parameter-settings}
\end{center}

\begin{center}
   \begin{tabular}{c c c}
        Number & $b$ & $p_\text{mig}$\\
        \hline
        6 & \texttt{3} & \texttt{0.2}  \\ 
        6 & \texttt{3.25} & \texttt{0.0, ... , 0.6}  \\ 
        6 & \texttt{3.5} & \texttt{0.0, ... , 0.6} \\
        6 & \texttt{3.75} & \texttt{0.0, ... , 0.6} \\
        6 & \texttt{4} & \texttt{0.2}  \\
        6 & \texttt{4.25} & \texttt{0.2}  \\
    \end{tabular} 
    \captionof{table}{\textbf{Runs of the abstract model}}
    \label{table:abstract-parameter-settings}
\end{center}

\subsection{Naturalistic Model} \label{appendix:naturalistic_model_parameters}

\begin{center}
   \begin{tabular}{c c p{10cm}}
        Category & Variable & Description  \\
        \hline
        \multirow{7}{*}{\textit{Game}} & $b$ & \texttt{55, 60, 65, 70, 75} \\
        & $c$ & \texttt{20} \\
        & $c_{\text{dist}}$ & \texttt{5} \\
        & $f$ & \texttt{20} \\
        & $c_{\text{alive}}$ & \texttt{2} \\ 
        & $c_{\text{repro}}$ & \texttt{2} \\ 
        \hline
        \multirow{6}{*}{\textit{Strategies}} 
        & $\epsilon$ & \texttt{1/20} \\
        & $\beta$ & \texttt{3/10} \\
        & $l$ & \texttt{1/20} \\
        & $d$, $d_{\text{mut}}$ & \texttt{NI}, \texttt{non-NI} \\
        \hline 
        \multirow{6}{*}{\textit{Environment}}
        & $p_{\text{mut}}$ & \texttt{1/100} \\
        & $p_{\text{mig}}$ & \texttt{0, 0.005, 0.05, 0.1, 0.2, 0.3, 0.4, 0.5, 0.6, 0.7, 0.8} \\
        & $s$ & \texttt{10} \\
        & $n$ & \texttt{20} \\
        & $g$ & \texttt{10} \\
        & $y$ & \texttt{20000}
    \end{tabular}
    \captionof{table}{\textbf{Parameters of the naturalistic model} *strictly speaking, these quantities are not the final costs and benefits; they are first divided by the number of agents foraging on the square} \label{table:naturalistic-parameter-settings}
\end{center}
Note that the migration rate is much lower in this model than the abstract one: this is because a generation in this round was around 41 rounds instead of just 10/3 rounds. We selected the migration rate so that lifetime migration rates would be equal between the two models.

\begin{center}
   \begin{tabular}{c c l}
        Number & $b$ & $p_\text{mig}$\\
        \hline
        8 & \texttt{50} & \texttt{0.5}  \\ 
        8 & \texttt{55} & \texttt{0.5}  \\ 
        16 & \texttt{60} & \texttt{0.0, 0.1, ... , 0.8} \\
        16 & \texttt{65} & \texttt{0.0, 0.1, ... 0.4, 0.6, ... , 0.8} \\
        8 & \texttt{65} & \texttt{0.005, 0.05} \\
        24 & \texttt{65} & \texttt{0.5} \\
        16 & \texttt{70} & \texttt{0.0, 0.1, ... , 0.8}  \\
        8 & \texttt{75} & \texttt{0.5}  \\
    \end{tabular} 
    \captionof{table}{\textbf{Runs of the abstract model}}
    \label{table:abstract-parameter-settings}
\end{center}
\section{Counterpart figures} \label{appendix:counterpart-figures} 

\begin{figure}[H]
    \centering
    \includegraphics[width=7cm]{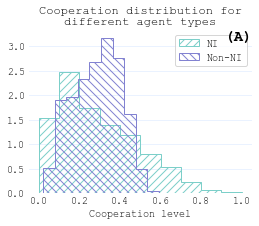}
    \caption{\textbf{Norm internalizer cooperation histogram.} The equivalent of Fig.~\ref{fig:norm-internalizer-quick-stats} for the naturalistic model. The norm internalizers have similarly higher variance in their coopereation levels.}
    \label{fig:norm-internalizer-quick-stats-naturalistic}
\end{figure}

\begin{center}
    \begin{figure}[H]
    \centering
    \begin{subfigure}{.5\textwidth}
        \centering
        \includegraphics[width=3cm]{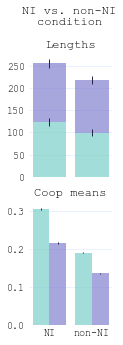}
    \end{subfigure}%
    \begin{subfigure}{.5\textwidth}
      \centering
      \includegraphics[width=3cm]{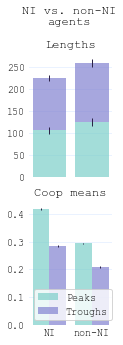}
    \end{subfigure}
    \caption{\textbf{Spike-trough comparison for the abstract model.} Equivalent of Fig.~\ref{fig:spike-summary} for the abstract model. It exhibits a similar pattern, except that the NI troughs are higher than the non-NI troughs (this is not the case in the naturalistic model)}
    \label{fig:abstract-spike-summary}
\end{figure}
\end{center}
\begin{center}
    \begin{figure}[H]
        \centering
        \centerline{\includegraphics[width=20cm]{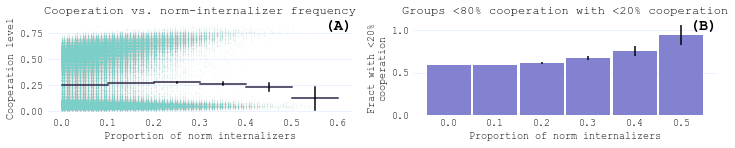}}
    \caption{\textbf{Polarization due to norm internalizers in the abstract model.} Equivalent of Fig.~\ref{fig:polarize} for the abstract model, exhibiting a similar pattern}
    \label{fig:abstract-polarize}
\end{figure}
\end{center}

\begin{center}
    \begin{figure}[H]
    \centering
    \includegraphics[width=13cm]{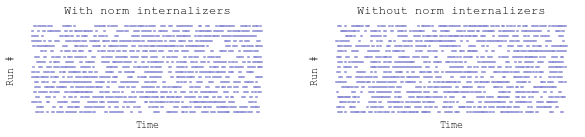}
    \includegraphics[width=13cm]{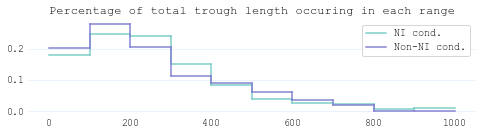}
    \includegraphics[width=13cm]{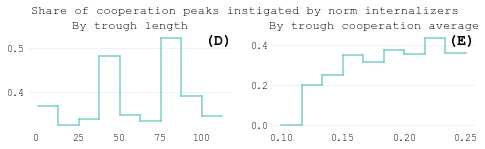}
    \caption{\textbf{Troughs characteristics in the abstract model.} These are the equivalents of Fig.~\ref{fig:spike-trough-comparison} for the abtract model. The abstract model does not exhibit the same pattern where norm internalizers prevent very long troughs}
    \label{fig:abstract-spike-trough-comparison}
\end{figure}
\end{center}

\begin{center}
    \begin{figure}[H]
        \centering
        \centerline{\includegraphics[width=16cm]{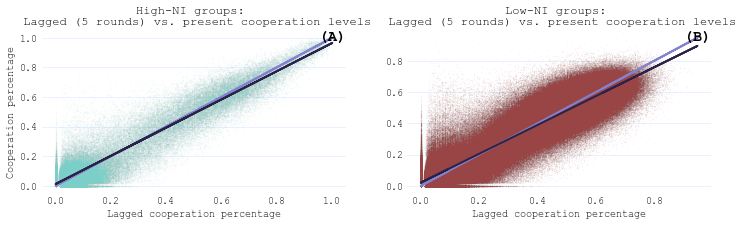}}
    \caption{\textbf{Norm internalizers as stabilizers in the naturalistic model.} Equivalent of Fig.~\ref{fig:lin-reg-stabilizers} for the naturalistic model, a similar pattern exists, with the coefficient for the low-NI state 0.92 ($\pm$ 0.0005), compared to the high-NI coefficient 0.95 ($\pm$ 0.0015). Confidence intervals are 95\% confidence}
    \label{fig:naturalistic-lin-reg-stabilizers}
\end{figure}
\end{center}

\begin{center}
    \begin{figure}[H]
    \centering
    \begin{subfigure}{.5\textwidth}
        \centering
        \raisebox{6mm}{
        \includegraphics[width=7.5cm]{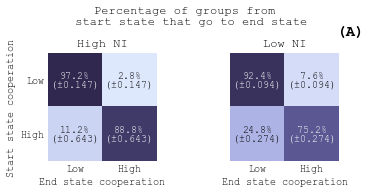}}
    \end{subfigure}%
    \begin{subfigure}{.5\textwidth}
      \centering
      \includegraphics[width=7.5cm]{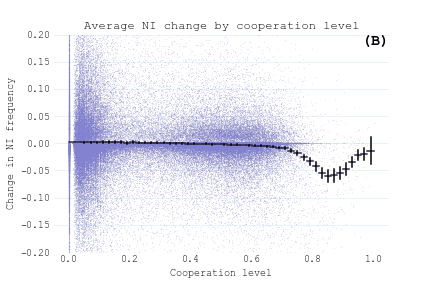}
    \end{subfigure}
    \caption{\textbf{Norm internalizers (NIs) as stabilizers, and their flaw - naturalistic model} The equivalent of Fig.~\ref{fig:stabilizer-and-flaw} for the naturalistic model, exhibiting a very similar pattern. The one difference is the uptick in norm internalizer change for very high cooperation levels --- this is because, at this level, there are very few free riders in the group. Groups in the abstract model never attained this level of cooperation}
    \label{fig:stabilizer-and-flaw-naturalistic}
\end{figure}
\end{center}

\end{document}